\documentclass[a4paper,fleqn]{cas-dc}

\usepackage[numbers]{natbib}
\usepackage{adjustbox}
\usepackage[dvipsnames]{xcolor}

\usepackage{xspace, flushend}
\usepackage{subcaption,breakurl}
\usepackage[normalem]{ulem}
\usepackage{tikz,algorithm,algorithmicx,algpseudocode,balance}

\def\tsc#1{\csdef{#1}{\textsc{\lowercase{#1}}\xspace}}
\tsc{WGM}
\tsc{QE}
\newcommand{\casaba}{{Caiti}\xspace} 

\newcommand*\mininumbercircled[1]{\tikz[baseline=(char.base)]{
		\node[fill=darkgray, text=white, shape=circle,draw,inner sep=0.1pt] (char) {#1};}}

\usepackage{wrapfig,graphicx}

\begin{document}
\let\WriteBookmarks\relax
\def\floatpagepagefraction{1}
\def\textpagefraction{.001}

\shorttitle{I/O Transit Caching for PMem-based Block Device}    

\shortauthors{Qing Xu, Qisheng Jiang, Chundong Wang}  

\title [mode = title]{I/O Transit Caching for PMem-based Block Device}

\author[]{Qing Xu}
\author[]{Qisheng Jiang}
\author[]{Chundong Wang}[orcid=0000-0001-9069-2650]
\cormark[1]
\ead{(cd_wang@outlook.com)}

\cortext[1]{Corresponding author at: School of Information Science and Technology, ShanghaiTech University, 393 Middle Huaxia  Road, Pudong, Shanghai, China 201210.}
\address[]{School of Information Science and Technology, ShanghaiTech University, Shanghai, China}

\begin{abstract}
	Byte-addressable non-volatile memory (NVM) 
	sitting	on the memory bus is employed to make {\it persistent memory} (PMem)
	in 
	general-purpose computing systems and embedded systems 
	for data storage. 
	{Researchers develop software drivers such as
		the block translation table (BTT) 
		to build block devices on PMem, so 
		programmers 
		can keep using mature and reliable conventional storage stack 
		while expecting high performance by exploiting fast PMem.} 
	However, our quantitative study shows that BTT underutilizes PMem and yields inferior
	performance, 
	due to the absence of the imperative in-device cache. 	
	We add a conventional I/O staging cache made of  DRAM space to BTT.
	As DRAM and PMem have comparable access latency, I/O staging cache is likely
	to be fully filled over time. 
	Continual
	cache evictions and {\tt fsync}s thus
	cause on-demand flushes with 
	severe stalls, such that 
	the I/O staging cache is concretely
	unappealing
	for PMem-based block devices.
	We  accordingly
	propose an algorithm named 
	\casaba with novel {\em I/O transit caching}. 
	\casaba eagerly evicts buffered data to PMem through CPU's multi-cores. 
	It also 
	conditionally bypasses a full cache and directly writes data into PMem
	to further alleviate I/O stalls.
	Experiments confirm  that 
	\casaba	significantly boosts {the performance with BTT} by up to 3.6$\times$, {without loss of block-level write atomicity.}
	
\end{abstract}


\maketitle

\section{Introduction}\label{sec:intro}
Non-volatile memory (NVM) technologies 
bring  about the availability of {\em persistent memory} (PMem) 
that is placed on the memory bus alongside DRAM, for CPU to load and store data.
As NVM 
generally has shorter access latency and higher write endurance
than flash memory, {researchers have considered using it for data storage in
general-purpose computing systems, cloud and virtual machines (VMs), internet-of-things (IoT) endpoints, and broad embedded systems}~\cite{lctes:Intermittent-Computing,NVM:cloud-MS,NVM:cloud-intel,NVM:vm-vmware}.
{Although Intel has discontinued its
	Optane DC memory business}~\cite{news:optane-dead,FS:SPFS:FAST-2023}, {the exploration of NVM technologies continues.
	The other types of NVM,
	such as STT-RAM}~\cite{NVM:Everspin-STT-MRAM}, {are still being developed 
	and deployed in embedded systems.}
{A number of new file systems for PMem have been developed}~\cite{FS:PMFS:Eurosys-2014,FS:HiNFS:tos,CHEN201818:UMFS,YANG2022102629:VSM,FS:SplitFS:SOSP-2019}.
These file systems mainly follow the direct access (DAX) fashion
to directly write and read files with PMem, bypassing DRAM page cache of operating system (OS).
In spite of exploiting the performance potential of PMem, they fail to satisfy 
a few fundamental requirements raised by applications with respect to reliability, compatibility,  deployment cost, and so on. 
{One of them is {\em block-level write atomicity}, which means that writing
	a block (e.g., 512B or 4KB) of data shall be done in an atomic (all-or-nothing) manner.}
{Typical applications and system softwares, such databases, rely on the atomic write of a block, because non-atomic write might leave a mix of up-to-date and stale data in one block and in turn cause the breach of data integrity.} {However,
	many of PMem-oriented 
	file systems do not support this feature, while
	conventional block devices have it}~\cite{dev:X-FTL:SIGMOD-2013,FS:BtrFS,FS:Hoare:OSDI-2020,FS:PMFS:Eurosys-2014,FS:SplitFS:SOSP-2019,FS:NOVA:FAST-2016}.

{As of today,
	PMem-oriented 
	file systems 
	have not been merged into Linux kernel or employed in production environments,}
{mainly due to the issues of reliability, compatibility, and deployment cost concerned by applications.}
{By contrast,} conventional block-based
file system mounted on a 
block device {has been the de facto storage stack for decades.}
{To be compatible with this storage stack,}
researchers have
tried to build block devices on top of PMem, e.g., PMBD~\cite{PMem:PMBD:MSST-2014}.
Intel's developers, {regarding the need of block-level write atomicity with Optane  memory},  
proposed the {\em block translation table} (BTT)~\cite{PMem:BTT-intel}. 
BTT is actively maintained as a software driver
in Linux kernel ~\cite{PMem:BTT} for 
programmers to make block devices from raw PMem and
mount conventional file systems like Ext4 or XFS.
Compared to their DAX variants without block-level write atomicity
(e.g., Ext4{-DAX} and XFS-DAX) 
that show `{\em WARNING: use at your own risk}' upon being mounted, 
{Ext4 and XFS are of maturity, robustness, and reliability after undergoing evolutions} in past years.
Programmers can continue using their tactics tuned for conventional storage stack 
with block I/O ({\tt bio}) interfaces that BTT provides atop fast PMem.

{As a software driver, BTT employs a combination of   copy-on-write (CoW) and logging to 
	achieve block-level write atomicity. However,  the cost of doing so is non-trivial.}
{
	We have done 
	a comparative test with mature Ext4 on PMem formatted with BTT,  mature Ext4 with raw PMem, 
	and Ext4-DAX.}
{Ext4 on PMem formatted with BTT
	yields inferior performance.}
{For example, when
	randomly writing 64GB data with 4KB per I/O request, 
	Ext4 with BTT 
	spends 37.4\% and 16.6\% more time than Ext4 with raw PMem and Ext4-DAX, respectively.} 
{Whereas, neither of the latter two  guarantees the block-level write atomicity.}

{When using fast PMem, we aim to yield high performance like raw PMem or Ext4-DAX
	while 
	preserving the block-level write atomicity for critical system and application softwares.}
{To attain this aim, we thoroughly analyze 
	the source code of BTT and compare it against traditional block 
	devices. We find that a critical component is absent in it, i.e.,}
the internal device cache\footnote{{{We interchangeably refer to the cache inside a block device
		as {\em DRAM cache}, {\em internal cache}, or {\em device cache}. However, for distinguishing, we always
	refer to the kernel-space page cache of OS with the combinational terminology {\em page cache}, without calling it
	{\em DRAM cache} or {\em cache}.}}} 
	{commonly installed in hard disk drives (HDD) or solid state drives (SSD) for I/O staging and 
	acceleration}~\cite{SSD:DuraSSD:SIGMOD-2014,10.1145/2947658,SSD:UFS-cache:TCAD-2020,flash:FTL2:LCTES-2013,cache:DIDACache:TOS-2018}.
The reason behind the absence of device cache
is that BTT has been designed in line with 
DAX, by which data is directly
written and read with PMem.
Assuming that we  complement BTT with a device cache, 
	we could use it to absorb write requests. This is likely to
	conceal the performance overhead caused 
	by maintaining block-level write atomicity at the BTT driver
	and in turn promote the overall performance. Inspired by this potential gain,
	we consider adding and managing
	a DRAM cache 
	within BTT in the I/O staging fashion. 
	We   follow two common polices to manage the cache when the cache space is used up.
{One is a PMBD-like caching
	that flushes a batch of buffered blocks when the cache is filled to an extent (watermark), e.g., 70\% or  100\% full}~\cite{PMem:PMBD:MSST-2014}.
The other one   {{evicts}} the least-recently-used (LRU) buffered block to make space.
We expect them to improve {the performance with BTT}.
However, with the foregoing test,
both 
algorithms 
decrease the performance  by 6.0\% and 15.1\%, respectively.
When no free space is left in the cache,
I/O requests have to 
stall for {{the drain 
of buffered data in one or multiple cache slots}}. 
{{Moreover, the upper-level file system such as Ext4 periodically (five seconds by default) 
		issues a {\tt bio} request with a
		 flag named {\tt REQ\_PREFLUSH} being set to asynchronously flush the internal cache of  block
		 device}}~\cite{footnote:write-back,footnote:five-sec}.
Frequent
on-demand flushes are hence occurring over time.
{An explicit call of {\tt fsync} is another source of cache flushes}~\cite{fsync,SSD:barrier:FAST-2018}.
Our further test confirms that I/O stalls also emerge when 
I/O staging caches serve {\tt fsync}s.

{The PMBD-like and  LRU caching policies represent the conventional 
	I/O staging strategy, which, however, is  practically ineffectual for BTT-like PMem-based block device with fast access speed}.
{Aforementioned tests motivate us to consider what a gainful device cache shall be for PMem managed by BTT driver.}
Firstly, it shall be simple and efficient 
to incur minimal performance penalty,
since PMem is relatively fast while 
{BTT itself 
	suffers from the software cost of preserving block-level write atomicity}.
Secondly, cache is likely to be fully filled at runtime.
Using {\tt fsync}s to flush data is also common
for applications.
We need to ensure that flushing buffered data does not cause stalls 
on the critical path of serving I/O requests.
Thirdly, as both DRAM and PMem are operated by CPU through the 
memory bus, we shall leverage multi-cores to handle I/O requests 
for high parallelism and efficiency.
Last but not the least, {as mentioned, the 
	caching strategy must not impair block-level write atomicity but simultaneously
	retain all features required by applications and file systems, such as the support
	for {\tt bio} flags and {\tt fsync}s}~\cite{fsync,SSD:barrier:FAST-2018}. 
{Taking into account these concerns, 
	we design a new 
	caching
	algorithm named \casaba (\textbf{\underline{ca}}ching with \textbf{\underline{I}}/O \textbf{\underline{t}}rans\textbf{\underline{i}}t) for PMem-based block device.
	The main ideas of \casaba are summarized as follows.}
\begin{itemize}
	\item  {\it \casaba eagerly evicts buffered data.}  
	Once the cache receives a data block,
	\casaba promptly launches a write-back for eviction, 
	instead of waiting for a flush or replacement.
	{It places buffered blocks in multi-queues and engages a pool of background
		threads in} {concurrently} {moving them to PMem.}
	\item  {\it  \casaba bypasses a fully filled cache.}
	When 
	no space is left in DRAM cache
	and a cache miss occurs at the arriving write request,
	\casaba directly writes data to PMem.{This avoids I/O 
		congestion 
		at the cache and further
		reduces response time on the critical path of serving I/O request.}
	\item {{{\it  \casaba exploits multi-core CPU for high
		concurrency}}} 	\\
	 {{{\it and scalability.}}} {{Each {\tt bio} 
	request carries an $lba$ and}} {{runs on a CPU core.
	\casaba logically partitions cache space into multiple sets
		and hashes each $lba$ to find an appropriate set without maintaining a mapping table.
		With a core handling a {\tt bio} request in one cache set, 
		\casaba manages to simultaneously proceed concurrent {\tt bio} requests.
	The multi-queues and background thread pool also
		accelerate concurrent write-backs of buffered data blocks to PMem and help to achieve scalability.}}
\end{itemize}

{In contrast to 
	I/O staging strategy intentionally buffers
	data for sufficiently long time,
	\casaba exploits scores of CPU cores to place data
	into cache in the foreground and swiftly \textit{transits} data to
	PMem in the background to avoid I/O stalls.}
Extensive experiments on benchmarks and applications show that
I/O transit caching enables \casaba to
boost {the performance with BTT} by up to 
3.6$\times$. {Besides PMBD and LRU}, {we further port and implement a state-of-the-art caching algorithm named Co-Active with PMem-based block device}~\cite{co-Active}.
{\casaba 
	significantly outperforms them with up to 3.6$\times$, 3.6$\times$, and 2.9$\times$ higher throughput, respectively.} 

The rest of this paper is organized as follows.
In Section~\ref{sec:bg} we briefly present PMem and BTT.
We show a motivational study in Section~\ref{sec:motivation}.
We detail the design of \casaba in Section~\ref{sec:design}
and evaluate it in Section~\ref{sec:eval}.
{We discuss related works in Section~\ref{sec:related}.}
We  conclude the paper
in Section~\ref{sec:conclusion}.

\section{Background}\label{sec:bg}

\subsection{The  Usage  of Persistent Memory}
PMem products are available in 
{{DRAM backed by flash memory} ({referred to as NVDIMM or NVDIMM-N)}}  
or NVM technologies such as 
spin-transfer torque RAM (STT-RAM)~\cite{NVM:Everspin-STT-MRAM}, 
{{phase-change memory (PCM)}}~\cite{6513577,10.1145/2627369.2627667,8715132,NVM:start-gap:Micro-2009,NVM:RBSG:IPDPS-2016},
and
{Intel Optane  memory}~\cite{Intel:optane}.  
{{PCM and Optane memory technologies generally hold inferior access speeds
compared to DRAM and are the main focus of this paper}}.
General-purpose computing systems, cloud and VMs, and embedded systems are using PMem
for 
storage~\cite{lctes:Intermittent-Computing,NVM:cloud-MS,NVM:cloud-intel,NVM:vm-vmware}.
As
CPU loads and stores data with PMem through the memory bus, 
researchers have 
developed new file systems upon using PMem as memory device
(e.g.,~\cite{FS:NOVA:FAST-2016,FS:SplitFS:SOSP-2019,FS:PMFS:Eurosys-2014}). 
These file systems mainly bypass OS's page cache with the DAX feature
and directly operate with files stored in PMem. 
{Without the involvement	of OS's page cache,
	DAX helps to reduce the cost of traversing software stack
	and alleviate the overhead of memory copying.
	With regards to space efficiency, the DAX feature saves 
	DRAM space that can be used for OS and applications.}

Though,
most of 
{file systems developed with DAX} are not merged in 
the mainline of Linux kernel. 
For Ext4- and XFS-DAX that are already contained
in Linux kernel because of being based 
on stable Ext4 and XFS, a mount of them shows a warning that says
`\textit{use at your own risk}'. 
To deploy them in product environments has 
to take into account
reliability, compatibility, and cost efficiency,
particularly regarding the absence of sector- to block-level (e.g., 512B or 4KB) write atomicity within them.
These require considerable efforts.
{In addition, the aforementioned Intel Optane memory
	was only one commercial NVM product shipped in a large capacity (up to  terabytes)
	to build scalable PMem. Whereas, Optane memory
	is physically
	not byte-addressable.
	Its access unit for write and read operations is 256B~\cite{PMem:study:FAST-2020}, 
	which  deviates from the assumption of DAX that PMem shall be accessible at the same byte granularity
	as DRAM. Using DAX on Optane memory-based PMem causes unnecessary write and read amplifications,
	which in turn hinder the use of    DAX   on real-world NVM products.}

Meanwhile,
how to 
use PMem
depends on the need of systems and applications.
{Mainstream file systems 
are built on the assumption of sector- to 
block-level atomic I/O that eases  
development, maintenance, and 
extension}~\cite{IO:atomic-BtrFS,FS:BtrFS,FS:Hoare:OSDI-2020,lctes:NBStack}. 
{Many applications,
	such as databases that have gained} {wide popularity} {for decades,
	explicitly or implicitly
	demand the support of 
	block-level write atomicity,
	because a mix of up-to-date and stale data in one block leaves ambiguity}~\cite{dev:X-FTL:SIGMOD-2013,FS:BtrFS}. 
As a result, {\bf configuring PMem in the form of block device
	is a promising and necessary alternative to use it}.

{{To facilitate the use of real PMem products, Intel provides two configuration modes. 
		One is {\em AppDirect} mode and the other one is {\em Memory} mode}}~\cite{PMem:study:FAST-2020,NVM:vmware-memory-mode}.
{{On one hand, the Memory mode uses PMem to expand main memory capacity without persistence}}~\cite{PMem:study:FAST-2020}.
{{When   PMem is configured in the Memory mode, it presents capacious but  volatile memory space, because the Memory mode employs DRAM as a cache to hide PMem’s higher latency, with hardware-controlled caching policy  between DRAM and PMem}}~\cite{Dash}.
{{On the other hand, if programmers intend to {\em explicitly} utilize a separate, visible PMem space, they must
		choose the {AppDirect} mode}}~\cite{PMem:study:FAST-2020,Dash,Viper}.
{{In the AppDirect mode, programmers can  
	create and use a {\em namespace} in different usage modes for different purposes}}~\cite{NVM:intel-btt-cmd-1,NVM:intel-btt-cmd-2,NVM:intel-btt-cmd-3}.
{{For example, they can set up a  namespace in the ``fsdax'' mode that exposes the raw PMem space without  sector- or block-level write atomicity}}~\cite{NVM:intel-btt-cmd-2}.
{{In practice, with such an exposed  PMem space, programmers can mount a file system with aforementioned DAX feature (e.g., Ext4-DAX), and they can also mount a mature file system (e.g., Ext4) in spite of no support for
	 block-level write atomicity}}.
{{Also, programmers can create a namespace for a ``sector'' mode and mount a mature file system with block-level write atomicity through the {\em block translation table}}}.

\subsection{Block Translation Table}
Researchers have developed software-based block devices on PMem, such as PMBD~\cite{PMem:PMBD:MSST-2014} and
	Block Translation Table (BTT)~\cite{PMem:BTT,NVM:intel-btt-cmd-3}.
	Both of them are kernel-space device drivers that achieve block-level write atomicity in the software approach.
As BTT has been actively maintained in Linux kernel since version 4.2, 
we focus on it in this paper.
{{In fact, the aforementioned ``sector'' mode for practically creating a PMem space with block-level write atomicity
		 is also known as ``btt'' mode or ``safe'' mode}}~\cite{NVM:intel-btt-cmd-1}.
{BTT is named after its main component, i.e., the block translation table recording the mapping from logical block number ($lba$) 
	to physical block number ($pba$). However, it is much more than address translation.} 
{To upper-level file system,}
BTT 
provides standard
{\tt bio} interfaces 
and allows programmers to configure the block size (e.g., 
4KB). 
Programmers thus can make and mount mature file systems such as Ext4 
on BTT.
{Inside the device,}
{it achieves 
	block-level write atomicity
	by a hybrid scheme of
	CoW and logging for data and metadata, respectively.}
{{In addition, programmers cannot mount a file system with DAX feature on a PMem space
		created in the ``btt'' mode to bypass the OS's page cache}}~\cite{NVM:intel-btt-cmd-1}.
{{However, they can still perform direct I/Os in the ``btt'' mode
		by using the O\_DIRECT flag that has been used  for conventional block devices such as HDDs and SSDs}}.	

\begin{figure}[t]
	\centering
	\includegraphics[width=1.0\columnwidth]{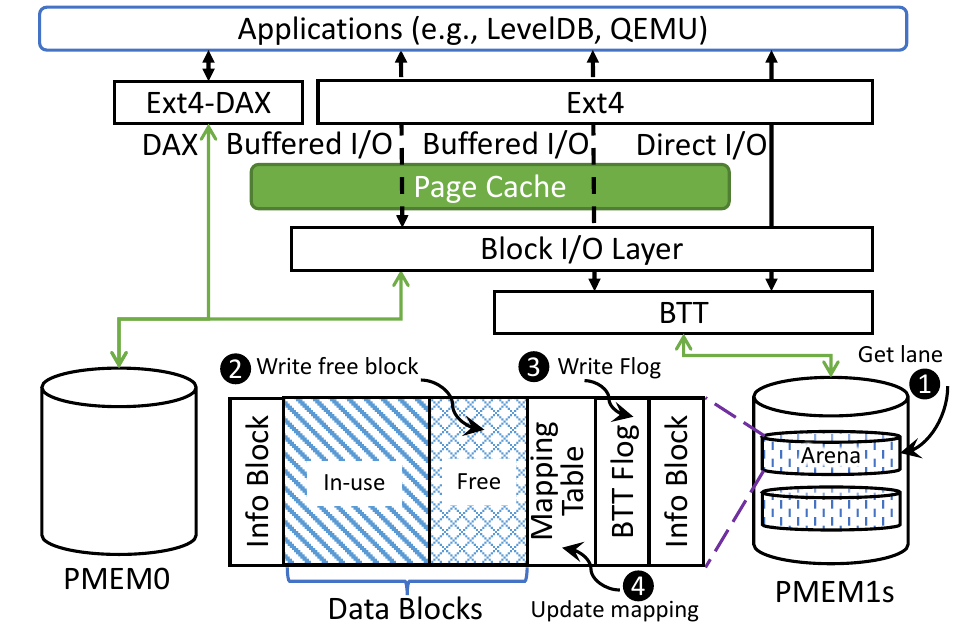}
	\caption{An Illustration of Block Translation Table (BTT)}
	\label{fig:btt}
\end{figure}

\autoref{fig:btt} illustrates the layout of {PMem space formatted with BTT and how BTT  manages  
	the PMem-based block device.} {BTT driver has no data or metadata kept in DRAM.} 
{It} divides the PMem space into multiple {\em arenas}. 
{Each} arena can have a maximum capacity of 512GB. Two identical
{\em Info blocks} are placed at the head and tail of each arena for redundant backup.
BTT divides an arena's PMem space
into {\em data blocks}, each of which is indexed with a $pba$. {As mentioned, BTT
	utilizes a  table to record address mapping from  $lba$s carried in {\tt bio} requests
	to $pba$s of PMem space.}

BTT introduces the concept of {\em lanes} 
to support simultaneous writes to PMem.
{The number of lanes is configured as 
	the number of CPU cores or 256, whichever is smaller.}
Among all data blocks per arena, 
BTT keeps a dynamic set of up to 256 {\em free blocks}.
Each lane is associated to a free block.
The {\em BTT Flog} of an arena is composed of multiple log entries,
used to track changes of address mapping 
from $lba$s to $pba$s when BTT is serving write requests.

{{Each {\tt bio} request is running on a CPU core with an ID} {for high concurrency}}.
This core ID determines in which lane BTT serves that {\tt bio} request.
On a read request with a target $lba$, 
{BTT driver locates the corresponding arena before going to the lane indicated 
	by core ID. It then looks up the mapping table and}
gets the $pba$. 
Next, BTT loads data indexed by $pba$ to complete the read request.
On a write request, as shown in~\autoref{fig:btt}, 
BTT still takes the 
{corresponding} lane (\mininumbercircled{1}).
{Instead of in-place updating, BTT writes data carried in the request
	into the lane's free block for out-of-place updating (\mininumbercircled{2}), i.e., the CoW way.
	Next, BTT initiates redo logging to record the change of address mapping for the $lba$.}
{BTT records 
	$lba$, old $pba$, new $pba$ of the lane's free block 
	in the log entry
	for crash recoverability}~(\mininumbercircled{3}).
After doing so, BTT modifies the mapping indexed by $lba$ in the mapping table
(\mininumbercircled{4}).
The previously-mapped block indexed by old $pba$ is factually swapped out to be free
and BTT employs this free block 
to supplement the lane.
{As a result, 
	each lane is always with an active free block  at runtime.} In summary,
CoW and logging jointly {ensure} block-level write atomicity for BTT with
the capability of rolling back 
on failed write for an $lba$ if 
system crashes (e.g., power outage or kernel panic).

\section{Motivation}\label{sec:motivation}

We have conducted a study on {a real PMem device} {{configured in the AppDirect mode}}
to observe the performance of BTT. More details of the platform 
are presented in Section \ref{sec:eval}.
We consider 
{three variants based on different usages of PMem.}
Firstly, 
we {{create a namespace in the ``btt'' mode and}
{mount Ext4 on the PMem-based block device}}. 
Secondly, we mount {mature}
Ext4 on bare-metal raw PMem {{created in the default ``fsdax'' mode}}. 
Thirdly, we mount Ext4-DAX on raw PMem {{reinitialized and created in the default ``fsdax'' mode}}.
They are denoted as {\tt BTT}, {\tt PMem}, and {\tt DAX}, respectively.
{As shown by the green arrows  in}~\autoref{fig:btt}{, both {\tt BTT} and {\tt PMem} utilize
	block I/O interfaces, while DAX bypasses the page cache of OS.}
Note that among them only {\tt BTT}  supports block-level write atomicity.
{\tt PMem} and {\tt DAX} 
may leave data corrupted if a power outage occurs. 

\begin{figure*}[t]
	\begin{subfigure}{0.38\columnwidth}
		\includegraphics[width=0.9\columnwidth, page=1]{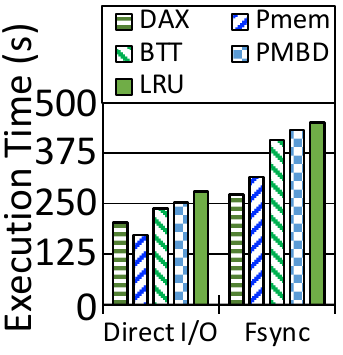}
		\caption{Execution time}\label{fig:mot:IO}
	\end{subfigure}
	\begin{subfigure}{0.62\columnwidth}
		\includegraphics[width=0.9\columnwidth, page=2]{./mot.pdf}
		\caption{{\tt fsync} time for {\tt PMBD}, {\tt LRU}, and \casaba}\label{fig:mot:fsync}
	\end{subfigure}
	\begin{subfigure}{\columnwidth}
		\includegraphics[width=0.9\columnwidth,page=1]{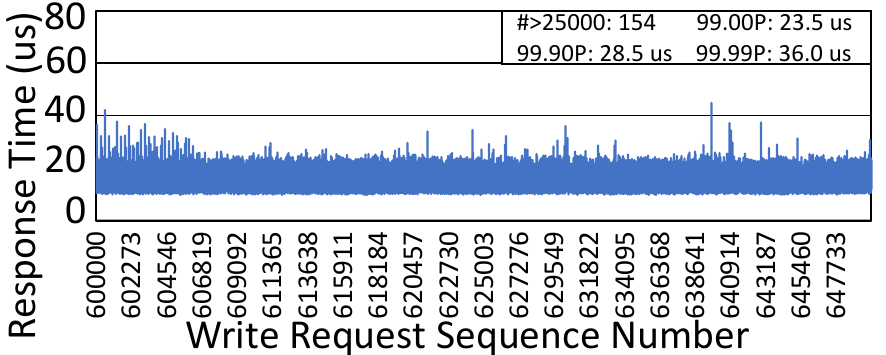}
		\caption{Runtime Response Time for {\tt BTT}}
		\label{fig:mot-curve-btt}
	\end{subfigure}
	\begin{subfigure}{\columnwidth}
		\includegraphics[width=0.9\columnwidth,page=2]{./curve-1-l.pdf}
		\caption{Runtime Response Time for {\tt PMBD}}
		\label{fig:mot-curve-pmbd}
	\end{subfigure}
	\begin{subfigure}{\columnwidth}
		\includegraphics[width=0.9\columnwidth,page=3]{./curve-1-l.pdf}
		\caption{Runtime Response Time for {\tt LRU}}
		\label{fig:mot-curve-lru}
	\end{subfigure}
	\caption{A Comparison on BTT, Ext4-DAX, PMem, and PMBD}\label{fig:mot}
\end{figure*}

To
quantitatively test 
three utilization ways,
we {{use Fio}}~\cite{benchmark:fio} {{to generate a write-intensive workload that}}
{{randomly 
writes a total volume of 64GB data in 4KB per I/O request  
under}} {{the direct I/O mode in order to exclude the impact 
of OS's page cache}}.  
~\autoref{fig:mot:IO} compares their results measured in execution time. 
Although  {\tt BTT} provides the strongest
crash consistency and 
write atomicity, it
spends 37.4\% and 16.6\% more time than
{\tt PMem} and {\tt DAX}, respectively. 
{\tt DAX} is   faster  than {\tt BTT} 
because Ext4-DAX, as 
tuned with the DAX feature,
has been optimized in the software stack for PMem.
BTT creates a block device
on raw PMem that is just used by {\tt PMem}. 
The 37.4\% gap between {\tt PMem} and {\tt BTT} 
indicates {{that, although {\tt BTT} has employed CPU's multi-cores to concurrently process I/O requests}},
 the cost of maintaining
block-level 
atomic writes by BTT driver.
In short,
{\bf {The performance with BTT} suffers from such non-trivial cost of enforcing block-level write atomicity}.

{When using fast PMem, we intend to make use of it for high performance
	like raw PMem or Ext4-DAX. In the meantime, we shall 
	retain the block-level
	write atomicity for critical system and application softwares.}
In order to improve {the performance with BTT}, we analyze the source {code} of it.
We find that
an internal device cache,
{a} key component that is widely built in modern 
HDDs and SSDs, is not contained in BTT. 
{I/O staging cache
is widely used to accelerate performance for storage}~\cite{SSD:DuraSSD:SIGMOD-2014,10.1145/2947658,SSD:UFS-cache:TCAD-2020,flash:FTL2:LCTES-2013,cache:DIDACache:TOS-2018}.
{For example, PMBD 
	employs a DRAM cache  to temporarily store dirty pages for I/O staging~\cite{PMem:PMBD:MSST-2014}. 
{{It has maintained}} a syncer daemon thread  
	that flushes the buffer to PMem when the buffer is filled to an extent (watermark). 
{{Though, the developers of PMBD did not fully consider
		how to utilize the multi-cores of CPU
		to accelerate caching for high concurrency when in designing
		PMBD, because they still followed
	the classic caching strategy used for a conventional block device. In short,}}	
	PMBD divides the {{DRAM cache}} into
	multiple sub-buffers. Once one sub-buffer is filled to the watermark (e.g., 70\% or 100\%), PMBD drains
	it by writing back buffered data to PMem.}
We have added multi-buffers in {{an overall capacity of 512MB}}
to BTT with PMBD-like caching (referred to
as {\tt PMBD}).
However,
in contrast to an expectation that I/O staging cache should have boosted the performance of BTT,
the execution time spent by {\tt PMBD}, as shown in~\autoref{fig:mot:IO},
is even 6.3\% {longer} 
than that of {\tt BTT}.
The flush of a sub-buffer upon preset watermark is likely to 
	cause high I/O congestion on the critical path. 
{{Moreover, Ext4 periodically performs a journal commit for crash consistency and data durability
	every five seconds by default, through issuing a {\tt bio} request with the {\tt REQ\_PREFLUSH} flag being 
set to  flush the volatile internal cache of storage device}}~\cite{footnote:write-back,footnote:five-sec}.	
{{However, Ext4 does not synchronously wait for the completion of such a periodical flush with the  {\tt SYNC} flag unset per request}}~\cite{203229,SSD:barrier:FAST-2018}.
{{Whereas, a user request that encounters one such asynchronous flush still
		suffers from additional overhead}}.
	 {{Besides {\tt PMBD}}}, {{we introduce
the conventional {\tt LRU} strategy that evicts 
the least recently used (LRU) cached block when the DRAM cache is full}}.
{{Yet {\tt LRU} still periodically flushes all data that it caches per
{\tt bio} request with {\tt REQ\_PREFLUSH} set}}.
As demonstrated by~\autoref{fig:mot:IO},
the execution time of {\tt LRU} is even a bit longer than that of {\tt PMBD}.

To figure out why {the performance with BTT} is not improved 
with the employment of I/O staging cache,
we have recorded the individual response time of serving every write request.
\autoref{fig:mot-curve-btt}, \autoref{fig:mot-curve-pmbd}, and \autoref{fig:mot-curve-lru} 
capture a part of these records in
a contiguous window of 50,000 requests 
for {\tt BTT}, {\tt PMBD}, and {\tt LRU}, respectively. 
The bottom margins in \autoref{fig:mot-curve-pmbd}  and \autoref{fig:mot-curve-lru}
are evidently narrower than that in \autoref{fig:mot-curve-btt}, which indicates
that writing data into cache truly 
helps to cause shorter response time ($\le$$5\mu$s) in serving a part of I/O requests.
However, a comparison among three diagrams clearly {shows} 
that
{\tt PMBD} and {\tt LRU} have completed
{numerous I/O requests 
	with more than 20$\mu$s, as depicted in}
\autoref{fig:mot-curve-pmbd} and \autoref{fig:mot-curve-lru}.
{Moreover, the continuous and frequent spikes exhibited in} \autoref{fig:mot-curve-pmbd} and \autoref{fig:mot-curve-lru}
{are not commonly present in} \autoref{fig:mot-curve-btt}.
At runtime, {\tt PMBD} is likely to use up DRAM space under I/O pressure 
and 
flushes buffered data to make space for incoming requests.
Similarly, {\tt LRU}  lays a 2-step  write
where {\tt LRU} firstly evicts a buffered block to PMem and then writes arriving data to DRAM cache.

\begin{figure*}[t]
	\begin{subfigure}{\columnwidth}
		\centering
		\includegraphics[width=0.95\columnwidth,page=2]{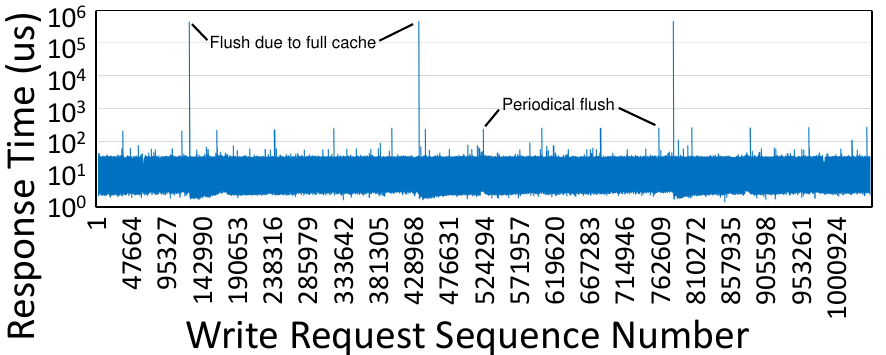}
		\caption{Runtime Response Time for {\tt PMBD} (512MB Cache)}
		\label{fig:mot-curve-pmbd-512}
	\end{subfigure}	
	\begin{subfigure}{\columnwidth}
		\centering
		\includegraphics[width=0.95\columnwidth,page=4]{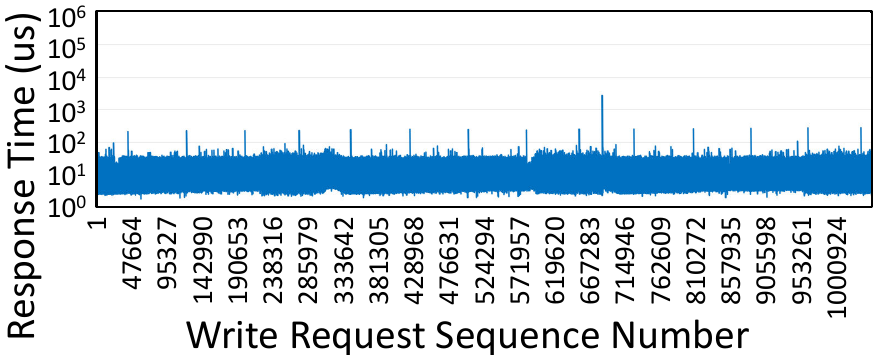}
		\caption{Runtime Response Time for {\tt LRU} (512MB Cache)}
		\label{fig:mot-curve-lru-512}
	\end{subfigure}	
	\caption{{{The response time for {\tt PMBD} and {\tt LRU} in a window of one million requests}}}\label{fig:curve-512-1mil}
\end{figure*}

{{Furthermore, the periodical flushes that Ext4 launches
	also cause {\tt PMBD} and {\tt LRU} to asynchronously flush   cached data.
To observe the impact of all flushes, we have captured
the runtime response time in a larger time window with one million (1,000,000) requests.
}}
\autoref{fig:curve-512-1mil} {{captures each request's response time for two caching algorithms. 
Both of them
exhibit almost uniformly distributed long tail latency over time.
Particularly,
in}} \autoref{fig:mot-curve-pmbd-512}, {{we can distinguish for {\tt PMBD} what tail latencies are caused by synchronous flushes (higher ones) due to a fulfilled
cache or periodical asynchronous flushes (relatively lower ones)  demanded by Ext4.
}}
{\bf {{Those flushes occur on the critical path
	of serving I/O requests, thereby generating frequent latency
	spikes}}}. 

In the meantime, applications explicitly call {\tt fsync}s
	to persistently store data and impose I/O ordering {{with the {\tt REQ\_PREFLUSH},  {\tt REQ\_FUA}, and {\tt SYNC} flags set}}, like what LevelDB does with 
	SSTable files for compactions~\cite{LSM:LevelDB}.
On receiving an {\tt fsync}, the device management {firmware or software} has 
to forcefully flush buffered data to persistent storage~\cite{SSD:barrier:FAST-2018}.
Since {\tt fsync}s are the other source of on-demand flushes,
we have modified the foregoing test case 
by inserting one {\tt fsync} after
every 128 write requests with 512KB (128$\times$4KB).
As shown by the right part of~\autoref{fig:mot:IO},
{\tt fsync}s evidently increase the execution time for all 
five. 
Note that 
{\tt PMBD} and {\tt LRU} still take 6.1\% and 10.8\% more time than {\tt BTT}, respectively.

In order to further explore how I/O staging cache influences the {\tt fsync} performance, 
we invoke an {\tt fsync} in different frequencies, 
i.e, after randomly writing 512KB 
to 128MB data. 
~\autoref{fig:mot:fsync} shows two close curves for
{\tt PMBD} and {\tt LRU},
respectively.
The {\tt fsync} time of them rises sharply 
as more data is written between two consecutive {\tt fsync}s. 
More  written data means more data is buffered in the cache.
As {\tt fsync} drains the cache, the impact of flushing
buffered data becomes more and more detrimental.
In practical, {\bf the cache is likely to be fully filled 
	over time
	since it continuously and uniformly receives data. On-demand flushes
	are hence inevitable and 
	severely impair performance of I/O staging cache}.

To sum up, {the performance with
	BTT 
	suffers from the cost of preserving block-level write atomicity}, while 
{the fashion of I/O staging 
	cache with frequent on-demand flushes
	cannot improve its performance.}
{{The  exploitation of CPU's multi-cores is also promising for caching data in a
PMem-based block device which is directly handled by the CPU}}.	
We thus 
reconsider how to
manage {an effectual} cache for BTT-like  block device made of fast PMem. 
The cache management must avoid flushing data
on the critical path of serving I/O requests.
It  shall be simple, without introducing dramatic 
performance penalty.
It may apply non-uniform caching policy under different conditions.
For example,
in case of a fully filled cache that is undergoing I/O congestion, 
it should handle arriving requests in a different but efficient way.
{Notably, BTT {\em emulates}
	a block device sitting on the memory bus. {{Both 
	DRAM and PMem are 
	operated
	by multi-core CPU. If we  {\bf exploit scores of CPU cores 
	to 	place data into cache in the foreground
	and swiftly transit data to PMem in the background}, 
we may effectively 
	boost the performance with BTT.}}   This summarizes the essence of \casaba.

\section{Design of \casaba}\label{sec:design}

\subsection{Overview}

\casaba manages a DRAM cache that
significantly differs 
from conventional I/O staging
caches. 
Firstly, 
it decouples cache space management from address mapping. 
By logically organizing cache space into sets and hashing $lba$s to localize a set, 
\casaba places a block freely in the cache and finds a block swiftly (Section \ref{sec:space}).
Secondly, in order to avoid 
stalls that prevent I/O requests from proceeding, 
\casaba has two writing policies to enable I/O transit caching (Section \ref{sec:wr}).
On one hand, \casaba   immediately initiates an {\em eager eviction} 
to write back
a block 
into PMem-based block device 
once the block is put down into a cache slot. 
On the other hand,
if no free space is available in the DRAM cache,
\casaba \textit{conditionally bypasses} the full 
cache but directly writes a block to PMem,
so as to circumvent   I/O congestion.
{\casaba bases both policies on CPU's multi-cores,
	which in turn enables scalability and concurrency for it.}
Thirdly, without losing block-level write atomicity,
\casaba comprehensively supports all standard {\tt bio} flags
and facilitates order preserving needed
by applications and system softwares (Section \ref{sec:discussion}).

{
	In the storage stack, \casaba's cache is positioned under the OS's page cache.
	They complement each other and share the functionality of
	buffering data for application and system softwares. The OS's page
	cache helps to absorb frequently accessed data for \casaba. In turn \casaba's cache helps to further reduce the time
	cost on the critical path of serving write requests.
	In addition, \casaba is software-only solution and incurs no change to
	hardware like memory controllers for PMem or DRAM. It has been practicable on a platform with real DRAM and PMem devices.
}

\begin{figure*}[t]
	\scalebox{1.0}{\includegraphics[width=\textwidth,page=1]{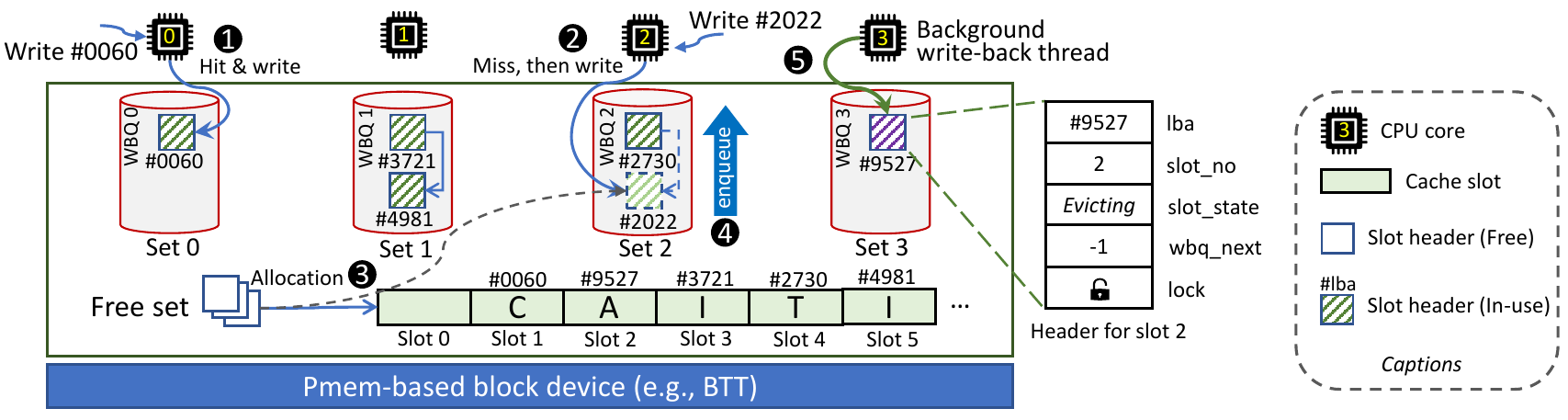}}
	\caption{Main Components and Write Procedure of \casaba}\label{fig:arch}
\end{figure*}

\subsection{\casaba's Cache Space Management}\label{sec:space}

\textbf{Organization.} 
{\casaba  demands and reserves 
	a contiguous DRAM space, e.g., in 512MB, from the OS.} 
It freely places blocks across this cache space and  partitions
the space into  cache
{\em slots}. For example,
the cache illustrated
in~\autoref{fig:arch}
has six slots.
All slots share a configurable uniform size, which
by default 
is
equivalent to the block size of BTT.
To track slots and handle I/O requests,
{\casaba logically maintains
	a number of cache {\em  sets} 
	holding valid data indexed by the hash value of $lba$.}
\casaba organizes each cache set as a write-back queue (WBQ).
As shown in~\autoref{fig:arch},
{\casaba also manages
	a global {\em free set} to group unoccupied slots over time}.

\textbf{Slot header.}
As illustrated by~\autoref{fig:arch},
\casaba tracks and manipulates a slot with {\em slot header}, in which
there are a slot number, an $lba$, a slot state, a WBQ pointer, and a lock. 
The lock is used for supporting concurrent accesses between multiple threads.
Slot number is the identity of a slot and
points to where a block is stored in the cache.
The $lba$ indicates where a block is stored in the PMem-based block device.
Although the mapping from an $lba$ to a slot number is fundamental for caching algorithms,
it only holds transiently for \casaba, because \casaba employs the I/O transit strategy
that swiftly sends a cached block down to device (see Section~\ref{sec:wr}).
Any slot staying in the free set is assigned an outlier $lba$, e.g., $-1$.
{Given a normal $lba$, \casaba hashes the $lba$
	to get the corresponding cache set number.
	The hash function exemplified in}~\autoref{fig:arch}
{is a modulo operation with four.} 
{In practical, we do a similar modulo operation 
	with $lba$ over the number of cache sets.}
{A collision may happen if two $lba$s are hashed to the same set.
	However, we allocate multiple cache slots per set and, more important,
	the eager eviction to be presented later helps to swiftly vacate cache slots. These jointly
	alleviate the impact of hash collisions.}

The WBQ pointer in each slot header is used to link slots in the cache set's WBQ 
as well as ones in the free set.
As to the slot state,
\casaba defines four legal ones, i.e., {\em Free},
{\em Pending}, {\em Valid}, and {\em Evicting}. 
Initially all slots are placed on standby in the free
set with the {\em Free} states. 
At the {\em Pending} state, 
data is being written
into the  slot. The 
{\em Valid} state
means  data
has been put in the slot while {\em Evicting} means
\casaba is doing write-back to PMem-based block device. 
After a successful write-back (eviction), \casaba recycles the slot to be {\em Free}
and puts it back to the free set for use in the near future.

\subsection{\casaba's Writing and Reading Policies}\label{sec:wr}

\subsubsection{\casaba's Writing Policies}\ \\ 
\indent The way \casaba handles a write request significantly differs from ordinary storage device caches.
Conventionally, since SRAM or DRAM embraces much shorter write latency than HDD and SSD,
researchers employ a 
cache to keep blocks buffered, expecting a sufficiently long period of I/O staging
to hide the slow access speed of underlying storage media.
When the device cache is full or an {\tt fsync} is received, the device management flushes 
partial or all buffered blocks for orderly persistence, which incurs drastically 
long response time
on the critical path. {Even though BTT builds a block device on top of relatively fast PMem, 
	it demands non-trivial cost to enforce block-level write atomicity} 
(see~\autoref{fig:mot:IO} and Section~\ref{sec:motivation}), so
waiting for data to be flushed and persisted is still costly for PMem-based block device.
As a result, we attempt to avoid the DRAM 
cache from being fully packed and minimize I/O stalls at runtime. 
This intention leads to two main policies \casaba utilizes in handling a write request.

\begin{algorithm}[t]	
	\caption{\casaba's Write Procedure (caiti\_write($lba$, $d$))}\label{algo:write}
	\begin{algorithmic}[1]
		\Require A write request that carries $lba$ and data $d$
		\State $\eta$ := {find\_set\_by\_hash}($lba$);\Comment{Hash to find cache set}\label{algo:line:hash}
		\State $\tilde{q}$ := get\_WBQ($\eta$); \Comment{Get the WBQ pointer $\tilde{q}$  of set $\eta$}	\label{algo:line:wbq1}				
		\State $sh$ := {dequeue}($\tilde{q}$, $lba$); \Comment{Get slot header $sh$ by dequeue}\label{algo:line:header}
		\If {($sh \neq$ {\bf NULL})}\Comment{Cache hit}\label{algo:line:hit}
		\State \Comment{We make \casaba first test and set the hit slot's state}
		\State test\_and\_set\_state($sh$, \textit{Pending});\Comment{state $\rightarrow$ \textit{Pending}}\label{algo:line:set-state}
		\State \Comment{\casaba writes data to the hit slot and set \textit{Valid} state}
		\State $ret$ := {write\_slot}($\eta$, $sh$, $lba$, $d$, \textit{Valid}); \Comment{state $\rightarrow$ \textit{Valid}} \label{algo:line:write-hit}
		\State $ret$ := {enqueue}($sh$, $\eta$, $\tilde{q}$); \Comment{\textit{\textcolor{cyan}{For eager eviction}}} \label{algo:line:enQ1}
		\Else \Comment{Cache miss, $lba$ is not matched in any slot}
		\If{({is\_cache\_full()} $==$ {\bf False})} \Comment{Cache is not full}\label{algo:line:not-full}
		\State 	\Comment{Let \casaba allocate a free slot $sh$ from the free set}
		\State $sh$ := {allocate\_slot\_from\_free\_set}($\eta$);\label{algo:line:allocate}
		\State set\_state($sh$, \textit{Pending}); \Comment{state $\rightarrow$\textit{Pending}} \label{algo:line:set-state2}
		\State\Comment{\casaba writes data  to slot $sh$ and sets \textit{Valid} state}
		\State $ret$ := {write\_slot}($\eta$, $sh$, $lba$, $d$, \textit{Valid});  \label{algo:line:write-new}	
		\State\Comment{Let \casaba get the WBQ of set $\eta$ for later eviction}		
		\State $\tilde{q}$ := get\_WBQ($\eta$); \label{algo:line:wbq2}	
		\State $ret$ := {enqueue}($sh$, $\eta$, $\tilde{q}$); \Comment{\textit{\textcolor{cyan}{For eager eviction}}} \label{algo:line:enQ2}							
		\Else \Comment{Cache is currently full}\label{algo:line:full}
		\State	$ret$ := {btt\_write}($lba$, $d$); \Comment{\textit{\textcolor{blue}{Conditional bypass}}}\label{algo:line:direct-write}
		\State \Return $ret$; \Comment{SUCCESS or -EIO}	\label{algo:line:direct-write-ret}			
		\EndIf
		\EndIf
		\If{($ret \neq$ {\bf False})} \Comment{Writing cache slot not failed}\label{algo:line:write-done}
		\State  $ret$ := notify\_eager\_eviction($sh$, $\eta$);\Comment{\textit{\textcolor{blue}{Eager Eviction}}}\label{algo:line:eager-eviction}	
		\EndIf	
		\State \Return $ret$; \Comment{SUCCESS or -EIO}\label{algo:line:normal-ret}								
	\end{algorithmic}
	\vspace{-1ex}
\end{algorithm}

\indent{\bf Eager eviction} is a policy that \casaba employs for prompt write-backs. 
In contrast to buffering a block for long-term I/O staging, 
\casaba immediately launches a write-back and delivers received blocks 
to underlying PMem-based   block device
through a background  thread.
By eager eviction, \casaba aims to free up cache slots in a timely fashion, so as not to engage incoming
write requests in
waiting for the completion of flushing buffered blocks 
due to no  available 
cache space. 
Concretely, \casaba manages to write arriving data to a free slot with one DRAM write
on the critical path, regardless of cache hit or miss, which, interestingly, 
renders the same effect of a write hit in I/O staging cache.
The latter even occupies cache space and affects serving other write requests.   
In addition, when receiving an explicit {\tt fsync} call that shall trigger the flush of device cache,
\casaba is able to quickly complete the  {\tt fsync}
since continuous eager evictions have drained most of the buffered data to PMem.

{\bf Conditional bypass} is the other writing policy of \casaba. 
Under high I/O pressure, it is possible that
no free slot is available in the DRAM cache.
{Waiting for a vacant slot by evicting a buffered block to make room would stall the current I/O request
	with a long response latency. 
	Whereas,
	\casaba bypasses the full cache and directly
	stores arriving block to  the PMem-based block device.} 
The reason why \casaba does so is twofold.
Firstly, a fully occupied cache means a tense I/O congestion is ongoing.
Waiting for a slot to be vacated by flushing the slot's buffered data down
further worsens the congestion.
Secondly, vacating a slot to make space
implies that the current write request cannot 
be finished until \casaba performs one PMem write 
and one DRAM write for putting down the arriving data into the vacated slot.
The conjunction of
these operations, however, costs even more time than a direct single write to PMem.

{\bf Write procedure.}
Eager eviction and conditional bypass jointly make \casaba function as
I/O transit rather than 
staging. Algorithm \autoref{algo:write} 
shows main steps
of \casaba's write procedure.
On receiving a {\tt bio} request with a target $lba$ and data,
it firstly hashes the $lba$ to decide the corresponding cache set (Line \autoref{algo:line:hash}).
Hashing helps to balance I/O loads among the entire cache space
as well as quickly identify a buffered block. The
hash  function
may incur conflicts with different $lba$s. As mentioned,
\casaba links the slot headers of  conflicting $lba$s in a cache set's WBQ. 
Over time,
the eager eviction policy of \casaba ensures that not many slots stay in one set.
\casaba gains access to the set's WBQ (Line \autoref{algo:line:wbq1}). It performs a 
scan 
to dequeue and grab the slot header for the target $lba$  (Line \autoref{algo:line:header}).
It expects a cache hit with a non-null slot header (Line \autoref{algo:line:hit}).
If so,
\casaba tests the slot's state (Line \autoref{algo:line:set-state}). Given a {\em Valid} state,
it transitions the slot to the {\em Pending} state
in order to prevent the slot from being written back
as the eager eviction is continually working.
{In case of the {\em Evicting} state, which means BTT is writing down the block to PMem, 
	\casaba 
	waits for the completion from BTT so as not
	to interrupt a persist operation. 
	This makes \casaba  retain block-level write atomicity.}
If \casaba finds the slot in the {\em Pending} state, which means a previous cache
write is still ongoing, it waits for the completion of that DRAM write.
Next \casaba updates data in the 
slot
and sets the slot's state as {\it Valid} (Line \autoref{algo:line:write-hit}).
It then
puts the slot 
to the set's WBQ for   eager eviction (Line \autoref{algo:line:enQ1}).

If a cache miss happens to target $lba$ 
while the cache is not full (Line \autoref{algo:line:not-full}),
\casaba allocates a free slot and places the slot into proper cache set $\eta$
(Line \autoref{algo:line:allocate}),
with the slot's state set as {\em Pending} (Line \autoref{algo:line:set-state2}).
Next,
\casaba puts down data into allocated slot for caching
and fills metadata 
in the slot's header, such as the $lba$ and {\em Valid} state (Line \autoref{algo:line:write-new}). 
Note that 
the state transition from {\em Pending} to {\em Valid}
is essential and useful for \casaba.
{During the duration of writing a block of data to DRAM, a read request that hits at the 
	cache slot with matched $lba$
	would not see incomplete data because of the {\em Pending} state.
	Also, \casaba would not do eager eviction on a slot unless the slot has accommodated 
	an integral block with a {\em Valid} state stationed.} 
Then,
\casaba fetches the entry pointer of WBQ (Line \autoref{algo:line:wbq2})
and prepares for eager eviction by enqueuing the slot (Line \autoref{algo:line:enQ2}).

In case of a full cache, \casaba directly persists the data 
to PMem-based block device and returns a success or not (Lines \autoref{algo:line:full} to \autoref{algo:line:direct-write-ret}). This corresponds to conditional bypass.
On the other hand,
every time a slot is placed in the set's WBQ,
\casaba
instantly notifies the submodule of eager eviction (Line~\autoref{algo:line:eager-eviction}).
Upon evicting a slot, \casaba changes the slot's state from {\em Valid} to {\em  Evicting}.
A
background thread now initiates the write-back. In the meantime,
\casaba returns a signal of success or fail for serving the write request (Line~\ref{algo:line:normal-ret}).

\autoref{fig:arch} briefly illustrates three cases for a write hit (\mininumbercircled{1}),
a write miss with a slot allocation (\mininumbercircled{2}\mininumbercircled{3}\mininumbercircled{4}),
and a concurrent write-back in the background for eager eviction (\mininumbercircled{5}).
As mentioned, the hash function used in \autoref{fig:arch} is the modulo operation with four.
On a write request with $lba$ {\it \#0060} that is executing on CPU core 0 (\mininumbercircled{1}),
\casaba hashes the $lba$ and gets set 0.
Then, through core 0, \casaba finds that a cache slot with the same $lba$ already stays in set 0's WBQ, 
thereby indicating a write hit.
Thus, \casaba 
writes data into the cache slot directly. 
After finishing the write, \casaba transitions the slot's state to 
{\it Valid}. 
As to the next write request with  $lba$ {\it \#2022} running on CPU core 2,
after hashing, \casaba does not find any slot holding {\it \#2022} among cache set 2.
Therefore, \casaba needs to handle this request as a write miss (\mininumbercircled{2}). 
Since the cache is not full, it allocates a free slot from the free set (\mininumbercircled{3}).
In the end, \casaba uses core 2 to set the slot as {\it Valid} as well and add it into set 2's WBQ
for eager eviction (\mininumbercircled{4}).
In \autoref{fig:arch}, \casaba is leveraging a background thread on core 3 to conduct eager eviction
(\mininumbercircled{5}). {\casaba finds a slot in the {\it Valid} state staying in set 3's WBQ.
	It transitions the slot state to   {\it Evicting} and starts writing back
	the data of $lba$ {\it \#9527}.
	In these procedures,
	read-write locks  of involved slots shall be acquired and released carefully in order to rule out 
	any disorders on writing and reading data for upper-level software layers.}

{\bf Scalable and concurrent I/Os.}
The efficiency of \casaba is supported by CPU's multi-cores.
An ongoing {\tt bio} request naturally 
executes 
on a CPU core that 
is responsible for proceeding the write to a cache slot or in-PMem location.
BTT functions as the backend of \casaba and embraces multiple   lanes that are ready to
take and handle multiple I/O streams simultaneously~\cite{PMem:BTT}.
As shown in ~\autoref{fig:arch}, 
\casaba maintains a thread pool. At runtime, it
grabs  
a background thread from the pool and runs the thread
on an idle CPU core. The  thread 
checks WBQs and writes buffered blocks within PMem-based block device.
In addition, 
\casaba leverages the lock 
in each slot header
to coordinate race conditions and
resolve write-write or write-read conflicts between concurrent
{\tt bio} requests on one $lba$.
{Additionally, the use of a global free set neither  posits a bottleneck for multi-threads
	nor impairs the scalability of \casaba.
	On one hand, when allocating or deallocating  a free cache slot,
	\casaba deals with the lock in the slot header through efficient {\em compare-and-swap} (CAS)
	operations. On the other hand, a global free set, rather than distributed local ones that are respectively
	dedicated to multiple CPU cores, is more friendly to workloads that are with write or read skewness.
	Assume that a workload  keeps writing data via few particular cores with multiple threads. Local free sets, if employed, 
	are likely
	to be used up for those working cores while the free sets of other cores stay idle.
	\casaba's globally shared free set has no such inefficiency in utilizing cache slots.}
{In summary,
	these above-mentioned components} concretely guarantee high scalability and concurrency 
for the collaboration between \casaba and BTT.

\subsubsection{\casaba's Reading Policy}\ \\ 
\indent{When \casaba receives a read request with an $lba$,
	\casaba firstly checks whether the $lba$ exists in a cache slot with {\em Valid} or {\em Evicting} state.
	This ensures that applications and file systems always perceive the latest valid and complete data.
	If a cache hit occurs, \casaba returns the buffered data copy from the matched slot.
	Otherwise, it redirects the read request to underlying PMem-based block device.} Additionally,
\casaba prioritizes write I/Os and a read miss in the DRAM cache
does not entail loading a block for buffering.

\subsection{Important Aspects of \casaba}\label{sec:discussion}

\textbf{Cache mapping and replacement.}
As mentioned, address mapping from storage space to cache space 
is basic for   cache management. {Employing a table for address mapping
was widely utilized in block devices}~\cite{10.1145/2512961}.  
Comparatively,
rather than
keeping
a mapping table,
\casaba performs the mapping from an $lba$ to a cache set by  hash  calculation. 
After hashing a target $lba$,
\casaba avoids
checking lots of slot headers in
the calculated cache set because of continual eager evictions in the background, so 
it
gains promising efficiency in satisfying on-demand I/O requests.
The conditional bypass also helps to
avoid cache replacement
for \casaba.
{Waiving the}  mapping table and replacement structurally distinguishes \casaba from
conventional caching designs.

\textbf{{\tt bio} and ordering.}
\casaba supports all {\tt bio} flags. 
{{The aforementioned {\tt REQ\_PREFLUSH} is used to flush the entire cache}}.
Applications such as databases and mail service frequently 
call {\tt fsync}s to orderly flush
file data to persistent storage. 
{An {\tt fsync} is eventually translated to  a {\tt bio} request with two flags, namely 
	{\tt REQ\_PREFLUSH} and {\tt REQ\_FUA}, turned on.
	{\tt REQ\_FUA}
ensures that storage device signals I/O completion only after the data has been persistently committed~\cite{footnote:write-back}. 
To support these flags for order preservation, 
\casaba flushes 
all  WBQ entries 
on receiving both {\tt REQ\_PREFLUSH} and {\tt REQ\_FUA}
and waits for I/O completion signals from underlying PMem-based block device. 
Then \casaba continues to serve subsequent {\tt bio} requests.
The alignment of {\tt bio} standard secures compatibility and viability for \casaba.
As to recent research works  that proposed
new methods to restructure emerging storage devices by adjusting store ordering
~\cite{SSD:barrier:FAST-2018,SSD:OPTR:ATC-2019,FS:Hoare:OSDI-2020},
\casaba can be easily adapted to suit their flags, primitives, or commands.

There is another ordering case in which an application 
	keeps writing the same $lba$ for consecutive updating.
	Given a high update frequency, the 
	OS's page cache is very likely to filter and absorb
	repeated write I/Os. As to 
	a low frequency, \casaba handles them as ordinary requests
	since previous data versions should have been evicted to PMem.
	Neither scenario affects \casaba's work, which in turn addresses its robustness.

{{A volatile internal cache has been widely employed for decades in  
HDDs and SSDs. 
The aforementioned periodical flush that Ext4 issues 
is used to ensure that, once a power fail or kernel panic happens,  at most
modified data put in the past five seconds would be lost in the volatile device cache}}~\cite{footnote:five-sec}.
{{If an application such as a database  demands a 
  higher guarantee of crash consistency and data durability, it shall  call
{\tt fsync} or {\tt fdatasync}, with the {\tt REQ\_PREFLUSH} set alongside write {\tt bio} requests to forcefully flush cached data}}~\cite{SSD:barrier:FAST-2018,footnote:write-back}.

{\bf Positioning.} 
\casaba is 
positioned in the latest developments of
software and hardware for storage stack.
For example, current
Linux kernel has included multi-queues ({\tt blk-mq}~\cite{footnote:blk-mq} for 
modern 
storage devices that BTT emulates with high parallelism.
\casaba's WBQs align with 
{\tt blk-mq}.
{Also, manufacturers 
	are shipping emerging block devices (e.g., Samsung CXL SSDs~\cite{footnote:CXL-SSD}) 
	with large 
	DRAM cache. The ideas of \casaba are applicable to
	enhance them.}

\textbf{Wear leveling.} 
Write endurance is crucial 
for
NVM~\cite{PMem:PCM:DAC-2012,NVM:curling:TCAD-2014,PMem:PCM:ICCAD-2019,HUANG2020101658,PMem:study:FAST-2020,PMSort,NVM:start-gap:Micro-2009,NVM:RBSG:IPDPS-2016} . 
Previous works have proposed wear leveling algorithms at both hardware and software levels.
{{For example,
researchers have reverse-engineered the hardware wear leveling
algorithm that Intel deploys for Optane memory}}~\cite{NVM:NVLeak:USENIX-2023,NVM:side-channel:USENIX-2023}.
{{On the other hand, software-based wear leveling is promising to prolong 
		the lifetime of PMem}} 
{{with wear-aware space allocation and recycling through the software stack}}~\cite{6513577,10.1145/2627369.2627667,8715132,227782,10.1145/3483839}.
{{During its procedures of eager eviction and conditional bypass, \casaba can take into account the
write endurance issue of NVM and collaborate with a 
software-based wear leveling scheme customized with the consideration of I/O transit caching. 
By doing so, \casaba is supposed to simultaneously gain both high performance and enhanced lifetime. We leave this as one of our  works for exploration in the near future.}}

\begin{figure*}[h]
	\centering
	\begin{subfigure}{\textwidth}
		\centering
		\scalebox{1.0}{\includegraphics[width=0.7\textwidth,page=1]{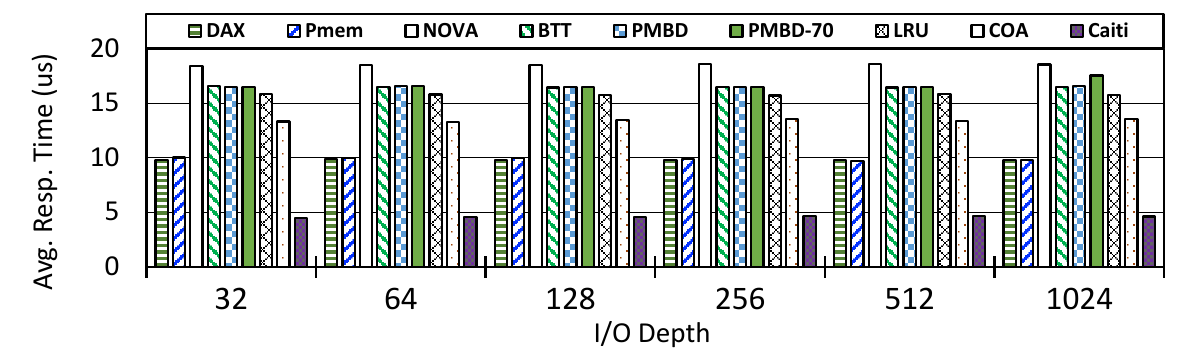}}
		\vspace{-1ex}
		\caption{Average Response Time}
		\label{fig:eval-fio-avg}
	\end{subfigure}
	\begin{subfigure}{\textwidth}
		\centering
		\scalebox{1.0}{\includegraphics[width=0.7\textwidth,page=4]{./fio729.pdf}}
		\vspace{-1ex}		
		\caption{Runtime Response Time for \casaba in  a window of 50,000 requests}
		\label{fig:eval-fio-curve}
	\end{subfigure}
	\begin{subfigure}{\textwidth}
		\centering
		\includegraphics[width=0.7\columnwidth,page=1]{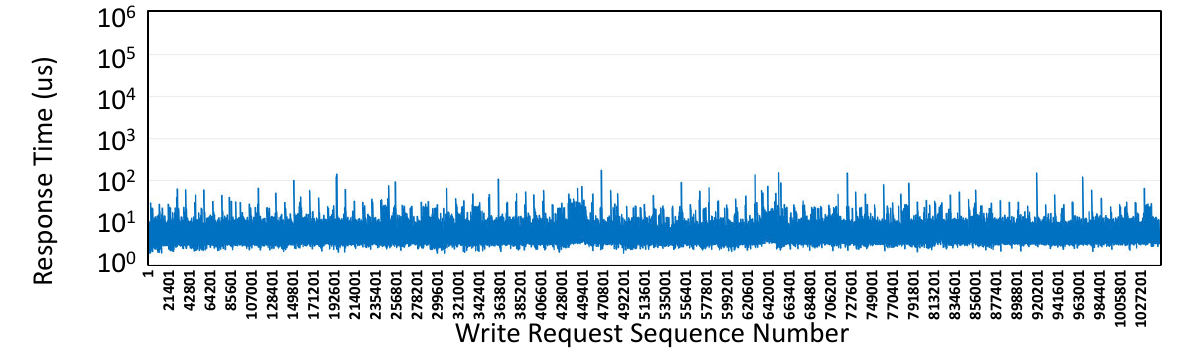}
		\caption{{{Runtime Response Time for \casaba in a window of one million requests}}}
		\label{fig:eval-curve-caiti-512}
	\end{subfigure}			
	\begin{subfigure}{\textwidth}
		\centering
		\scalebox{1.0}{\includegraphics[width=0.7\textwidth,page=2]{./fio729.pdf}}
		\vspace{-1ex}		
		\caption{99.99P  Latency (log scale)}
		\label{fig:eval-fio-99p}
	\end{subfigure}
	\begin{subfigure}{\textwidth}
		\centering
		\scalebox{1.0}{\includegraphics[width=0.7\textwidth,page=3]{./fio729.pdf}}
		\vspace{-1ex}		
		\caption{Multi-threading Test}
		\label{fig:eval-fio-threads}
	\end{subfigure}
	\caption{A Comparison with Fio on Average/Runtime Response Time, Tail Latency, and Multi-threads}\label{fig:eval-fio}
\end{figure*}

\section{Evaluation}\label{sec:eval}

\textbf{Setup.} 
{All experiments have been conducted on a machine with real PMem products, i.e., Intel Optane
	DC memory in 768GB configured in the  {\em AppDirect} mode. There are also 384GB DRAM installed in the machine. The CPU is 36-core Intel Xeon Gold 6240.
	The OS is Ubuntu 20.04.5 with Linux kernel 5.1. The compiler is GCC/G++ 8.4.0.
	The vanilla BTT has 2,001 lines of code (LOC).
	We add or change 1,066 LOC  
	to implement the core functions and structures of \casaba.}

{We compare BTT with \casaba to 
	Ext4-DAX, Ext4 mounted on raw PMem, original BTT,
	and NOVA in CoW mode}~\cite{FS:NOVA:FAST-2016}.
{{They represent common ways of using PMem}} 
{and are referred to as \texttt{DAX}, \texttt{PMem}, \texttt{BTT}, 
	and {\tt NOVA}, respectively.}
{Note that {\tt DAX} and {\tt PMem} do not provide block-level write 
	atomicty}.	
{As to caching with the I/O staging strategy, we consider 
	algorithms including aforementioned PMBD and
	LRU, as well as the state-of-the-art Co-Active}~\cite{co-Active}.
{For PMBD-like caching, we have implemented two variants.
	One is denoted as {\tt PMBD}.
	It
	flushes the entire cache if the cache is 100\% full (see Section~\ref{sec:motivation}).
	In other words,
	when there is no free cache slot available upon the arrival of a write request,
	{\tt PMBD} does a cache flush.  
	The other one, referred to as {\tt PMBD-70},
	strictly follows the literature and source code of PMBD.
	{\tt PMBD-70}
	flushes the buffer when the buffer is
	70\% full. {\tt PMBD-70} further differs from {\tt PMBD}
	in that the former flushes the cache through a syncer daemon  thread}~\cite{PMem:PMBD:MSST-2014}.
Unlike {\tt PMBD} {and {\tt PMBD-70}}, 
{\tt LRU} evicts the LRU 
slot rather than an entire cache if cache is full.{
	Co-Active is a collaborative active write-back cache management approach. 
	We port it from NVMe SSD to PMem-based block device managed by BTT.
	It employs a
	cold/hot separation module to distinguish cold and hot data via a Bloom Filter. Meanwhile, it maintains
	two linked lists for dirty and clean blocks in DRAM cache, respectively. 
	When PMem stays idle, 
	Co-Active proactively evicts data from the dirty list to PMem.
	We refer to it as {\tt COA}.}
	We configure the block size as 4KB for BTT.
	{{\casaba, {\tt PMBD}, {\tt PMBD-70}, {\tt LRU} and {\tt COA} are four designs that manage
		a DRAM cache on top of PMem-based block device
		and for each of them}}, we set the  cache slot size and
	default cache capacity as 4KB and 512MB, respectively.

We conduct experiments with micro-benchmarks (e.g., Fio)
and real-world applications (e.g., LevelDB~\cite{LSM:LevelDB}
and QEMU~\cite{qemu:cache})
to thoroughly evaluate \casaba.
{The main 
	metrics to measure performance 
	is response (execution) time and throughput (bandwidth in MB/s)}.

\subsection{Micro-benchmark (Fio)}\label{sec:eval-fio}

Fio is a powerful and capacious benchmark~\cite{benchmark:fio}.
{The reason why we choose it for evaluation is multifold.}
{Firstly, it can continuously generate I/O requests with varying I/O depth   to launch different pressure tests.
	Secondly, it reports meaningful results that help
	to observe and interpret the performance of \casaba. Thirdly, it allows us to configure multiple threads, enabling us to test the scalability of \casaba. 
	Fourthly, it further supports a configurable composition of I/O engine, access pattern, I/O depth, and multiple jobs (threads) to comprehensively evaluate caching algorithms under comparison.
}

With Fio,  we firstly test \casaba with substantial I/O pressure.
{{As we have done in the motivational study}},
{{we configure Fio in 
the direct I/O mode, with libaio engine continuously issuing random writes for 30 minutes
in a 64GB file 
with 4KB I/O size}}. We 
vary the I/O depth from 32 to 1024 
so that the storage stack 
undergoes increasingly more orderless I/O requests in flight~\cite{FS:Hoare:OSDI-2020}.
{By doing so, we aim to impose stressful workloads
	and thus deeply explore the capability of \casaba with ample I/O pressure}.
As shown in~\autoref{fig:eval-fio-avg}, 
\casaba consistently outperforms
other designs by taking much less response time.  For example,
with 128 I/O depth, it boosts {the performance with BTT}  by up to 3.6$\times$, while 
{\tt PMBD}, {{\tt PMBD-70}}, {\tt LRU}, {{\tt COA}} and {\tt NOVA} that enforce block-level write atomicity 
incur 3.6$\times$, {3.6$\times$},
3.5$\times$, 
{2.9$\times$} and 4.1$\times$ execution time compared to \casaba, respectively.
The superior performance of \casaba justifies the efficacy of its strategy of I/O transit caching
with
timeliness and concurrency.
On massive write requests arriving,
\casaba handles them with multiple cache sets supported by 
multi-cores
in the foreground
and 
leverages a pool of background threads and WBQs to promptly 
move buffered data to PMem.
As a result, I/O stall and congestion hardly occur to \casaba.
Meanwhile,
{\tt BTT} and {\tt NOVA}  directly write data to PMem with
efforts for 
achieving block-level write atomicity. {\tt PMBD}'s cache, albeit
employing multi-buffers, is likely to be overfilled at runtime and
stall 
for the drain of buffers. 
{Although {\tt PMBD-70} flushes the cache when the cache 
	space is used as much as 70\%, it cannot timely handle burst requests 
	with the syncer daemon  thread, resulting in limited improvement.}
{\tt LRU} also frequently stalls 
due to
its 2-step write upon no free cache space  (see Section~\ref{sec:motivation}) and yields similar performance compared to {\tt PMBD}. 
{Compared to other I/O staging algorithms, {\tt COA} reduces response 
	time by taking   advantage of cold/hot data separation. 
	However, {\tt COA} is still inferior to \casaba. 
	Upon the arrival of continuous I/O requests, PMem is unlikely to be
	idle. {\tt COA} has to evict buffered data and stall incoming requests. As a result,
	many cache replacements occur on the critical path for  {\tt COA}. 
	This is a common issue shared by caching algorithms with the
	I/O staging strategy.
	Separating cold and hot data also incurs cost for it}.
The performance gap {between these I/O staging algorithms and \casaba is  concretely 
	significant as regards I/O stall and congestion the formers are encountering}.

{\tt DAX} and {\tt PMem}
spend 115.7\% and 120.2\% more time than
\casaba, respectively.
They directly persist data to PMem, while
\casaba puts data to DRAM cache. 
{\casaba's hashing on $lba$s distributes data in multiple sets
	to leverage multi-cores and eases its eager
	eviction that concurrently vacates filled cache slots. Conditional bypasses
	also occasionally alleviate I/O congestion at DRAM cache.}
In addition,
Yang et al.~\cite{PMem:study:FAST-2020}
obtained a similar observation 
as they found 
that 
writing data to PMem has lower throughput than
doing so with DRAM.

Secondly,
we have captured the runtime
response time per request for \casaba. 
We present 50,000 points of response time in~\autoref{fig:eval-fio-curve}
and a comparison with~\autoref{fig:mot-curve-btt} to~\autoref{fig:mot-curve-lru} 
conveys that \casaba spends much shorter time in serving each request than 
	{\tt BTT}, {\tt PMBD}, and {\tt LRU}.
{{
At runtime, the number of flushes does not change significantly for \casaba,
since Ext4 consistently issues a {\tt bio} request with the {\tt REQ\_PREFLUSH} set
for flush every five seconds. Though, the eager eviction
has moved  data aggressively to PMem and results in much more lightweight flushes for \casaba compared
to {\tt PMBD} and {\tt LRU}}}. {{This explains why the majority of runtime latencies}}
in  \autoref{fig:eval-fio-curve} {{for \casaba is below 20$\mu$s}}. 
 {{In fact, we also record the runtime response time in the large window of one million requests and shown their points in}}
 \autoref{fig:eval-curve-caiti-512}.
 {{A comparison between}}  \autoref{fig:eval-curve-caiti-512} and \autoref{fig:curve-512-1mil} {{indicates that
 		in a long run, \casaba yields much shorter
 runtime response time  at the presence of periodical flushes 
 since eager eviction and conditional bypass help to reduce the volume of cached data 
that is supposed to be flushed every five seconds}}.
{{Though, foreground and background threads of \casaba may contend for the same cache slot
		and incur lock/unlock operations. Meanwhile,
		 many processes are running simultaneously and numerous 
		events (e.g., I/O interrupts or exceptions) are concurrently happening. When serving
		an I/O request, the CPU core on which \casaba is running might be scheduled
		to run for the other process or handle a sudden event.  
These two factors are likely to bring latency spikes to \casaba over time.}}
To grab a more quantitative and comprehensive view on the critical path of serving write requests,
we
have recorded the 99.99 percentile (99.99P) tail latency that Fio reports after finishing all  requests.
The tail latencies shown in~\autoref{fig:eval-fio-99p} with Y axis being in the logarithmic scale
complement our observation with~\autoref{fig:eval-fio-avg}. 
{{\casaba generally achieves shorter tail latency by I/O transit caching.}}
With greater I/O depth, all nine algorithms experience higher pressure
	and the tail latency 
	dramatically increases.
\casaba is likely to trigger more conditional bypasses to avoid I/O congestion
while, for instance, {\tt PMBD} 
at 1024 I/O depth makes 14.0$\times$ 99.99P latency  than \casaba
because {\tt PMBD} insists on waiting for the flush of buffered data.

Thirdly, 
{we test the scalability of \casaba by varying the number of Fio's jobs (threads)
	under 32 I/O depth}.
In~\autoref{fig:eval-fio-threads},
with more   jobs (1 to 32) 
issuing write requests, 
\casaba is  more performant than
others. 
{It  
	works with multiple sets by dynamically utilizing multi-cores
	over its foreground and background
	threads. \casaba thus 
	yields high performance in a scalable and balanced way
	with multi-threading workloads.}
{However, {\tt COA} shows a clear distinction 
	between a single job and multiple jobs. This is because
	{\tt COA} has difficulty in handling
	concurrent and continuous I/O requests from multiple threads,
	since its
	cold/hot data
	separation approach cannot efficiently identify overwhelming data. 
}

Fourthly, we observe the impact of cache capacity on
{\casaba, {\tt PMBD}, {\tt PMBD-70}, {\tt LRU} and {\tt COA} with one job and 32 I/O depth.}
{We set six cache sizes which, as shown in Table}~\ref{tab:cache-size},
hardly affect performance for all. 
This phenomenon is mainly due to the large volume of data received at the cache.
For example, 
in the starting first minute,
no less than 15GB data has been written.
With massive data continuously arriving, 
a cache in tens of or even more gigabytes stays overloaded.
This again explains why  I/O transit caching is effective and useful for PMem.

\begin{table}
	\caption{{{The Impact of Cache Size (Avg. Resp. Time in $\mu$s)}}}
	\begin{center}	
		\resizebox{\columnwidth}{!}{
			\begin{tabular}{|r|r|r|r|r|r|r|}
				\hline		 	 	 	 	 	 	 	 	 	  	 
				Capacity & 64MB  & 128MB  & 256MB   & 512MB   & 1GB  & 2GB    \\ \hline	\hline
				{\tt PMBD} & 	16.81	& 16.50	& 16.67 &	16.39 &	16.53	& 16.23 		\\ \hline	
				{\tt PMBD-70} & 	16.84	& 16.75	& 16.58 &	16.47 &	16.47	& 16.29 		\\ \hline
				{\tt LRU}	         &	 	16.72&	16.95&	16.42&	16.85&	16.55 &	16.53  \\ \hline
				{\tt COA}	         &	 	12.89&	12.81&	12.85&	12.95&	12.55 &	12.30  \\ \hline
				\casaba	         &		4.44&	4.42&	4.29&	4.41&	4.44 &	4.37  \\ \hline
		\end{tabular}}\label{tab:cache-size}
	\end{center}
\end{table}

Fifthly,{we measure the spatial cost of metadata \casaba and other algorithms need for caching.} 
{For every 4KB block (cache slot), the spatial cost for each caching algorithm is as follows.}
\begin{itemize}
	\item {\casaba with 102B in all: 8B for $lba$,  4B for slot\_number,  1B for state,  40B for  lock,  33B for work\_struct, and 16B for two pointers (WBQ and free list).}
	\item {{\tt PMBD, PMBD-70} and {\tt LRU} with 84B in all:  8B for $lba$, 4B for slot\_number, 40B for  lock, and 32B for lists.}
	\item {{\tt COA} with 102B in all: 8B for $lba$, 4B for slot\_number, 40B for  lock, 48B for lists, and 2B for Bloom Filter.}
\end{itemize}
{For \casaba, 512MB cache costs about 12.75MB and
	a 4KB slot demands 102B on average.
	The ratio
	of 2.5\% ($\frac{102}{4096}$) indicates high space efficiency for \casaba.}

\subsection{The Breakdown for \casaba}

\begin{figure*}[t]
	\centering
	\begin{subfigure}{\columnwidth}
		\includegraphics[width=0.9\columnwidth,page=2]{./BD-119-1213.pdf}
		\caption{Time Breakdown of Calling {\tt pwrite}}
		\label{fig:eval-bd-write}
	\end{subfigure}
	\begin{subfigure}{\columnwidth}
		\includegraphics[width=0.9\columnwidth,page=1]{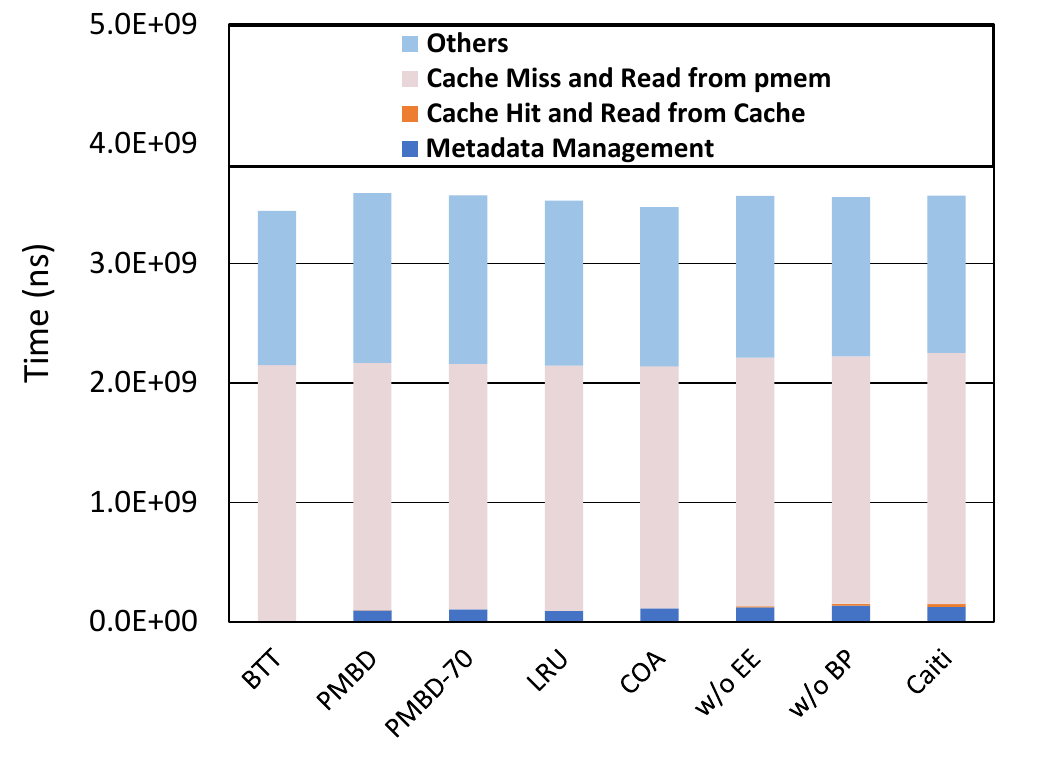}
		\caption{Time Breakdown of Calling {\tt pread}}
		\label{fig:eval-bd-read}
	\end{subfigure}
	\begin{subfigure}{\columnwidth}
		\includegraphics[width=0.9\columnwidth,page=3]{./BD-119-1213.pdf}
		\caption{The Percentages of Cache and Non-cache Writes}
		\label{fig:eval-bd-utilization}
	\end{subfigure}
	\begin{subfigure}{\columnwidth}
		\includegraphics[width=0.9\columnwidth,page=4]{./BD-119-1213.pdf}
		\caption{Average Time under Different Write Conditions}
		\label{fig:eval-bd-cond}
	\end{subfigure}
	\caption{{{A Detailed Breakdown of Different Cache Management Categories}}}\label{fig:eval-bd}
\end{figure*}

{In order to comprehensively investigate what factors contribute to the 
	performance of \casaba and other caching algorithms, we have conducted a breakdown test.
	To obtain relatively pure time trajectory, 
	we choose the most fundamental POSIX I/O calls for the test.
	We overhaul the critical paths of calling POSIX write and read ({\tt pwrite} and {\tt pread}), respectively,
	when each caching algorithm serves them. The main steps 
	of \casaba include
	`cache
	metadata management', `cache eviction and write' (a stalled write), `conditional bypass', `cache write only' (due to cache hit or a vacant slot), `WBQ enqueue', `cache flush'
	({due to Ext4 committing its journal every five seconds}~\cite{footnote:five-sec} or {\tt fsync}~\cite{fsync}), 
	and others (e.g., software overhead across kernel- and user-spaces).
	{{Moreover, in order to measure the respective impact of eager eviction or conditional bypass, 
	we also include two variants of \casaba with either eager eviction or conditional bypass disabled. They are denoted
as `w/o EE' and `w/o BP', respectively}}.
	In our test program, we
	continuously send write requests to a file stored in PMem.
	{{The file is opened in the O\_DIRECT mode
	to rule out the impact of OS's page cache.}}
	Each write operation is performed at a granularity of 4KB.
	Overall	
	1024$\times$1024 write operations are done. The target offset in the file for each write
	is randomly generated in advance by following a uniform distribution.
	The contributions of aforementioned factors are captured in
	\mbox{~\autoref{fig:eval-bd-write}}. 
	With these results, we have obtained following key insights.
}

	Firstly, the occurrence of `cache eviction and write' is extremely rare for \casaba, which means that almost no stall
	has been detected for \casaba. This is because
	the eager eviction timely makes space  
	through multiple background threads and an arriving write request
	does not need to wait for a free cache slot. As shown by} \autoref{fig:eval-bd-write}, {{\tt PMBD}, {\tt PMBD-70}, {\tt LRU}, and {\tt COA}
	all suffer from stalls caused by insufficient cache space
	and spend 40.5\%, 25.3\%, 40.0\%, and 32.7\%
	on `cache eviction and write', respectively.
	Comparatively, the time taken for \casaba's
	all writes on the critical path of {\tt pwrite}, including
	both `cache write only' and `cache eviction and write',
	is just 11.5\% of all the time cost.

{{Secondly, without either eager eviction or conditional bypass, the time cost spent on corresponding
	operations dramatically rises up and impacts the overall performance of \casaba. 
	For example, as shown in}}~\autoref{fig:eval-bd-write},
{{with the bar of `w/o EE', the absence of eager eviction accumulates data in the DRAM
	cache and conditional bypasses are triggered more frequently, thereby incurring more percentage
	of conditional bypasses. In the meanwhile, without conditional bypass, all data blocks
	must be written into cache slots, which helps to increase cache hits (i.e., `Cache Write Only').
	Though, the {\tt pwrite} workload studied here is quite simple 
	with regard to simply writing 1024$\times$1024 data blocks and I/O stalls hardly happen. 
	Consequently,
	with both eager eviction and conditional bypass enabled, \casaba achieves close performance to the variant with
	eager eviction only.}}

{{{Thirdly}}, as to the critical path for {\tt pread}, each caching algorithm
	achieves a similar breakdown (see} \autoref{fig:eval-bd-read}). 
{We note that our test program writes and reads
	data across a 4GB space, which is eight times the cache capacity of
	512MB. Meanwhile, in the Linux kernel, Ext4 commits its journal every five seconds (periodical write-back)
	and in turn enforces a cache flush with {\tt REQ\_PREFLUSH} flag}~\cite{footnote:write-back}.
{Consequently, the limited cache capacity and periodical cache flushes jointly entail
	cache misses.
	Yet the penalty of loading target data from PMem
	is equivalent among all caching algorithms.}

{{{Fourthly,}} \autoref{fig:eval-bd-utilization}
illustrates the percentages of `writing cache only' due to cache hits or available free slots, `cache eviction and write', as well as `conditional bypass' over all writes.
	It is evident that \casaba almost handles all write requests by writing data
	to cache slots. This again justifies why \casaba yields superior write performance.
	The syncer daemon  thread helps {\tt PMBD-70} to vacate cache slots
	and in turn exhibits different percentages on `cache eviction and write'
	compared with {\tt PMBD}.
	In addition,
	{\tt COA} writes more data directly to cache than {\tt LRU} and {\tt PMBD}.
	However, the more complex management of {\tt COA} demands 
	substantially more processing time, especially upon a {\tt REQ\_PREFLUSH} flag
	due to Ext4's periodical write-back every five seconds.
	As shown in \autoref{fig:eval-bd-write}, {{\tt COA}
	spends 1.9$\times$ time compared to that of {\tt LRU} and {\tt PMBD}
	on `cache flush'.
	{{Comparatively, \casaba eagerly evicts cached data} {to underlying PMem-based block device through multiple concurrent threads in the background}}. 
	{{As mentioned, upon a periodical flush every five seconds, \casaba is very likely to write down 
 only a handful of data to PMem}}.
	{{This in turn justifies why the time \casaba uses to	flush the cache is almost negligible as shown in}}
	\autoref{fig:eval-bd-write}.

	{{Fifthly}}, \mbox{\autoref{fig:eval-bd-cond}} displays the average time needed for `writing cache only', 
	`cache eviction and write', and `conditional bypass'. 
	For {\tt PMBD-70}, the time for `cache eviction and write' increases due to contention
	between the daemon thread and the working thread (e.g., for the list lock).
	Conversely, \casaba's `conditional bypass' costs shorter time than other algorithms'
	`cache eviction and write', highlighting the efficiency of conditional bypass 
	when the cache is under congestion.

	{{Sixthly}}, as told by the bars in \autoref{fig:eval-bd-write},
{the remaining time (`others') accounts for a large 
	proportion in the overall time cost. To gain a deep insight, we capture  the detailed timeline 
	of 100 consecutive write requests and present a contiguous period in \mbox{\autoref{rev:eval-bd-caiti}} for both \casaba and {\tt BTT}. In the upper half,
	we mark the timestamps when the test program has initiated the 
	write requests, when these requests reach \casaba, when \casaba places data into 
	a WBQ, when \casaba completes processing each write request,
	and when \casaba's background thread finishes evicting the cached block. 
	The lower half of \mbox{\autoref{rev:eval-bd-caiti}} is for original {\tt BTT}.
	The latency between a request's issuance and
	the request's arrival at \casaba 
	is significant, taking 54.0\% of the time cost per write request on average.
	This latency is mainly caused by the software penalty due to handling a system call between user- and kernel-spaces}~\cite{DBLP:books/wi/SGG2018,280870,10.1145/2150976.2151017,10.5555/3358807.3358858,9373910}.
{The latency, however, provides a sufficient time window for \casaba's background threads
	to eagerly evict cached data to PMem.
	Consequently, given that a program consecutively writes data 
	to the same file location twice   without the involvement of OS's page cache, 
	it is impossible that \casaba's eager eviction
	stalls the second write request from being served since 
	the cache slot would be timely vacated when the request is still traversing the software
	stack. In the buffered I/O mode, the OS's page cache would
	accumulate two consecutive file writes before they are sent to underlying device driver.
}

{Last but not the least, the metadata management accounts for a very small 
	proportion in the overall time cost for \casaba. As shown by the bottom segment in} \autoref{fig:eval-bd-write},
{it only takes 2.9\% for \casaba. This low percentage addresses the cost efficiency of \casaba's management tactics.
	Take slot allocation/deallocation for example.
	Allocating or deallocating a free slot with the  free set
	is done through efficient CAS operation. Furthermore,
	as \casaba makes the free set 
	globally shared, 
	each core is able to quickly fetch
	a vacant slot and avoids being blocked upon serving highly-skewed workloads.
}

\begin{figure*}[t]
	\centering
	\scalebox{1.0}{\includegraphics[width=0.6\textwidth,page=2]{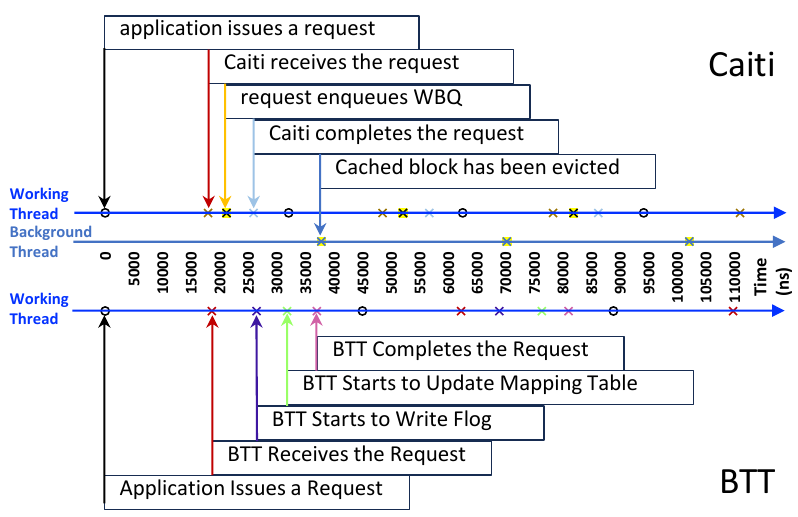}}
	\caption{Runtime Time Fragments of {\tt pwrite} for \casaba and BTT}
	\label{rev:eval-bd-caiti}
\end{figure*}

\subsection{Real-world Application}
\subsubsection{LevelDB}

\begin{figure*}[t]
	\centering
	\begin{subfigure}{\textwidth}
		\centering
		\scalebox{1.0}{\includegraphics[width=0.7\textwidth,page=1]{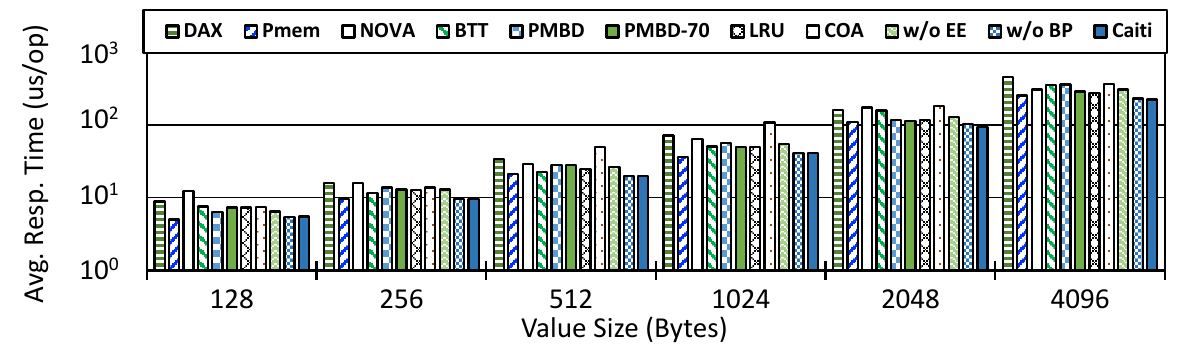}}
		\caption{Fillrandom (log scale)}
		\label{fig:eval-leveldb-fillrandom}
	\end{subfigure}
	\begin{subfigure}{\textwidth}
		\centering
		\scalebox{1.0}{\includegraphics[width=0.7\textwidth,page=2]{./leveldb-1212.pdf}}
		\caption{Overwrite (log scale)}
		\label{fig:eval-leveldb-overwrite}
	\end{subfigure}
	\begin{subfigure}{\textwidth}
		\centering
		\scalebox{1.0}{\includegraphics[width=0.7\textwidth,page=3]{./leveldb-1212.pdf}}
		\caption{Readrandom}
		\label{fig:eval-leveldb-readrandom}
	\end{subfigure}
	\begin{subfigure}{\textwidth}
		\centering
		\scalebox{1.0}{\includegraphics[width=0.7\textwidth,page=4]{./leveldb-1212.pdf}}
		\caption{Readhot}
		\label{fig:eval-leveldb-readhot}
	\end{subfigure}
	\caption{{{A Comparison with LevelDB on Fillrandom, Overwrite, Readrandom, and Readhot}}}\label{fig:eval-leveldb}
\end{figure*}

We illustrate with 
LevelDB for several 
reasons. 
Firstly, LevelDB is a 
key-value store 
gaining wide popularity. 
Secondly, LevelDB frequently calls {\tt fsync}s to 
persistently store data through storage stack and 
causes on-demand flushes not covered in Fio tests.
Thirdly, Fio performs random I/Os
in block size 
while
LevelDB generates bulky I/Os batched in megabytes
with SSTable files.  
Fourthly, {{we run Fio under direct I/O mode but with LevelDB, 
OS's page cache and LevelDB's application-level
caches 
are involved.}}
Fifthly,  {{\tt db\_bench} built in LevelDB
	has various write- and read-intensive workloads.}
While we focus on optimizing write performance, 
we
fully test \casaba's capability  
in handling read requests with LevelDB.

~\autoref{fig:eval-leveldb} shows the average response time 
of 
all designs 
on serving ten million requests issued by 
{\tt db\_bench}
on fillrandom, overwrite, readrandom, and readhot
with  value size varied from {128B to 4KB}. 
Take 2KB values for example.
\casaba spends 40.6\% and 38.2\% less time than
{\tt BTT} on fillrandom and overwrite, respectively.
Still with 2KB values and fillrandom, \casaba takes 41.9\%,  46.2\%, 20.0\%, {17.8\%}, 19.6\% and {48.8\%} less time 
than {\tt DAX}, 
{\tt NOVA}, {\tt PMBD}, {{\tt PMBD-70}}, {\tt LRU} and {{\tt COA}}, respectively. 
LevelDB writes an SSTable in 4MB or 2MB 
with an {\tt fsync} followed.
Once receiving such a bulk of data, \casaba
absorbs with cache slots
and background threads launch eager evictions.
Ideally, when a foreground thread finishes writing a  slot,
a background thread promptly initiates write-back.
When {\tt fsync} drains the cache, 
\casaba leaves only a handful of slots being in {\em Valid} states.
If many 
SSTables arrive and congest 
at the cache,
\casaba  forwards them to PMem for avoidance of stalling.
Comparatively, 
{{\tt PMBD}, {\tt PMBD-70}, and {\tt LRU} 
	forcefully drain buffered
	data upon {\tt fsync} or full cache} while {\tt DAX}'s
and {\tt NOVA}'s DAX does not use OS's page cache for buffering. 
{
	As to {\tt PMBD-70}, overwhelming bulky I/Os continually fill the 70\% cache capacity 
	and many cache evictions thus have to happen on the critical path.
	{\tt COA} struggles to find an opportune moment to move out dirty data
	to PMem.  However, the cold/hot prediction strategy is rendered ineffective 
	for   {\tt COA},
	{since all SSTable files are written only once to become immutable}}.  
{Worse,  {\tt fsync}s continually drain the cache over time and hinder {\tt COA}
	from accurately identifying cold or hot data. 
}
{For the same reasons, the
	{\tt fsync} time of \casaba  is also much lower than that of {\tt PMBD} and {\tt LRU}, as shown in}~\autoref{fig:mot:fsync}.  
{\tt PMem} yet achieves comparable
or a bit higher write performance
because it benefits from no cost for block-level write atomicity and
the use of OS's page cache for I/O staging.
~\autoref{fig:level:dis}
depicts the cumulative distribution 
of response time for LevelDB on fillrandom with 4KB values. 
Evidently \casaba climbs to the end of 100 percent 
at a much steeper rate.
A comparison on these curves
also tells  that \casaba incurs shorter tail latency, thereby enabling 
higher quality of service.

As to read-intensive workloads,
both 
readrandom and readhot 
follow a uniform distribution to fetch data{,}
except that readhot's accesses are limited {to} a small range 
to {simulate the} 
hot spots commonly found in real-world situations. 
A comparison between 
~\autoref{fig:eval-leveldb-readrandom}
and~\autoref{fig:eval-leveldb-readhot}
shows that 
readhot generally demands much less response time.
This is because it is easier for LevelDB's and OS's caches 
to  buffer and hit hot data than all data. 
{\tt DAX} and {\tt NOVA} are two that read data with DAX bypassing OS's page cache, so 
they spend increasingly more time in loading larger values from PMem.
By taking advantage of
upper-level
caches,
\casaba yields comparable read performance than others,
which further justifies the soundness of its 
write buffering strategy.

{{In addition, we also use aforementioned four workloads issued by db\_bench
to test the  two variants with either eager eviction or conditional bypass disabled, i.e., `w/o EE' and `w/o BP', respectively.
As shown in}}~\autoref{fig:eval-leveldb-fillrandom} {{and}} \autoref{fig:eval-leveldb-overwrite}
{{with two write-intensive workloads, the full \casaba spends less time
than either variant lacking one of two components.
Interestingly, as shown in}}~\autoref{fig:eval-leveldb-readrandom} {{and}} \autoref{fig:eval-leveldb-readhot}
{{for two read-intensive workloads, i.e., readrandom and readhot, respectively,
the variant without eager eviction generally achieves a bit shorter average response time
than  two other \casaba variants with eager eviction. The reason is that, the eager eviction strategy aggressively
evicts cached data down to PMem-based block device and is likely to result in more cache misses, leading to longer response latency.
}}

\begin{figure*}[t]
	\centering
	\begin{subfigure}{\columnwidth}
		\vspace{-3ex}
		\includegraphics[width=\columnwidth]{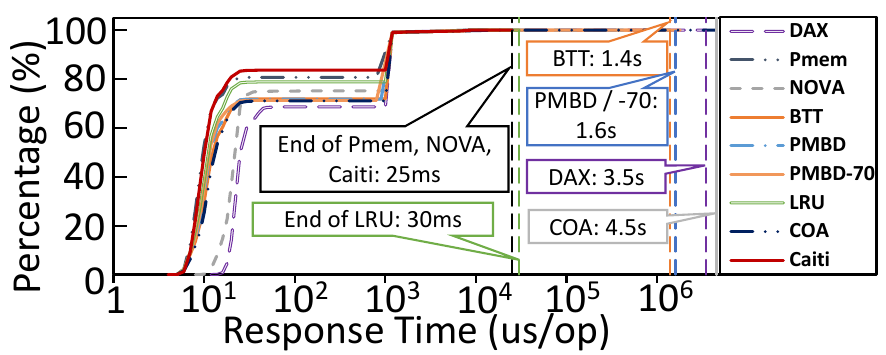}	
		\caption{A Cumulative Distribution of Response Time for LevelDB on Fillrandom with 4KB  Values}
		\label{fig:level:dis}
	\end{subfigure}
	\begin{subfigure}{\columnwidth}
		\includegraphics[width=1.0\columnwidth,page=1]{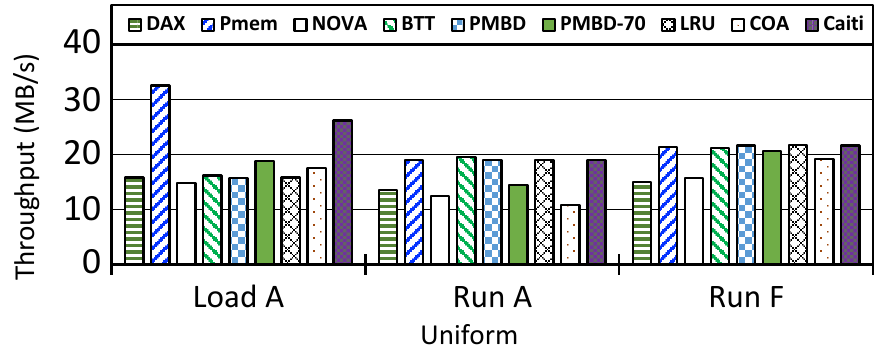}
		\vspace{-1ex}
		\caption{YCSB Benchmark with Uniform Distribution on LevelDB}
		\label{fig:eval-leveldb-ycsbu}
	\end{subfigure}
	\begin{subfigure}{\columnwidth}
		\includegraphics[width=1.0\columnwidth,page=2]{./ycsb.pdf}
		\caption{YCSB Benchmark with Zipfian Distribution on LevelDB}
		\label{fig:eval-leveldb-ycsbz}
	\end{subfigure}
	\begin{subfigure}{\columnwidth}
		\includegraphics[width=1.0\columnwidth,page=3]{./ycsb.pdf}
		\caption{YCSB Benchmark with Latest Distribution on LevelDB}
		\label{fig:eval-leveldb-ycsbl}
	\end{subfigure}
	\caption{Fillrandom Latency and YCSB Throughput on LevelDB }
	\label{fig:eval-leveldb-ycsb}
\end{figure*}

{To showcase the versatility of Caiti in a more rigorous manner,
	we subject LevelDB to typical workloads generated by YCSB
	(Yahoo! Cloud Serving Benchmark)~\cite{YCSB-CITE}. Besides loading
	data into LevelDB, we evaluate \casaba with YCSB's A  
	(update-heavy workload with 50\% point queries and 50\% updates) and F 
	(workload with 50\% read-modify-write and 50\% read) that are both mixed
	with write and read requests. We configure uniform, zipfian,
	and latest distributions to cover various data access patterns that mimic
	real-world  scenarios.}
{
	As shown by} \autoref{fig:eval-leveldb-ycsbu} to \autoref{fig:eval-leveldb-ycsbl},
{\casaba still achieves higher or comparable performance than other caching algorithms.
	For example,
	as depicted in \mbox{~\autoref{fig:eval-leveldb-ycsbz}} with data following
	the skewed Zipfian distribution, the throughput of
	\casaba is 65.8\%, 40.0\%, 66.3\% and 51.8\% higher
	than {\tt PMBD}, {\tt PMBD-70}, {\tt LRU} and 
	{\tt COA}, respectively, 
	upon loading data. 
	\casaba also outperforms other caching algorithms under the latest distribution with
	workloads A and F which alternately write and read data over time. We note that the
	highly-skewed latest distribution
	always chooses the most recent data for operation. As indicated by} \autoref{fig:eval-leveldb-ycsbl}, {\casaba efficiently exploits the locality of
	cached data. These results with varying data access distributions justify the rationality of Caiti's I/O transit strategy with eager eviction and conditional bypass.
}

\begin{figure*}[t]
	\centering
	\begin{subfigure}{\columnwidth}		
		\includegraphics[width=1.0\columnwidth,page=1]{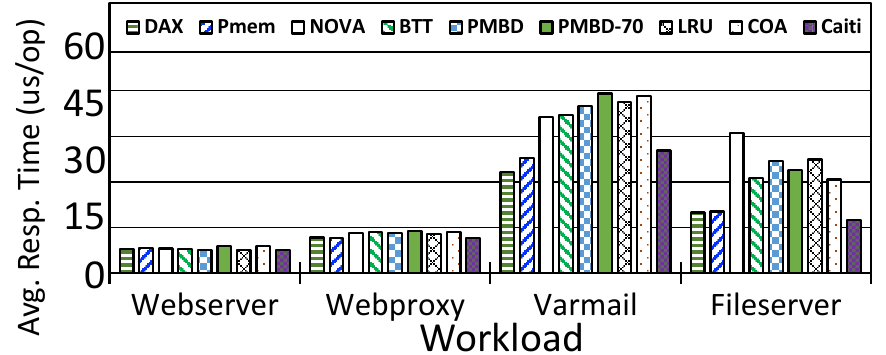}
		\caption{Comparison of Overall Performance with Filebench}\label{fig:qemu}
	\end{subfigure}
	\begin{subfigure}{\columnwidth}		
		\includegraphics[width=1.0\columnwidth,page=2]{./qemu801.pdf}	
		\caption{Comparison of Read Performance with Filebench }\label{fig:qemu-read}
	\end{subfigure}
	\caption{A Comparison with Filebench on Webserver, Webproxy, Varmail, and Fileserver in VM }
	\label{fig:eval-qemu}
\end{figure*}

\subsubsection{Virtual Machine}

Virtual machines (VMs) are
widely 
used in multi-tenant cloud 
and practitioners have
considered deploying PMem to enhance VMs~\cite{NVM:cloud-MS,NVM:cloud-intel,NVM:vm-vmware}. 
To evaluate the efficacy of \casaba for VM, 
we utilize Filebench~\cite{benchmark:filebench} to
generate different access patterns with   Webserver, Webproxy, Varmail, and Fileserver workloads in a VM with disk images atop PMem.
The guest OS 
is Ubuntu 21.04 with Ext4 as file system,
running within QEMU 6.2.0 and KVM~\cite{qemu:cache,qemu:kvm}. 
VM disk images are set in the RAW format with virtio enabled. 
We use the default writeback cache policy for QEMU
and the VM is with 32GB DRAM and eight vCPUs.

~\autoref{fig:qemu} depicts average response time for
\casaba and other algorithms with
aforementioned workloads 
running in VM. 
\casaba reduces the average response time by 12.0\%, 13.3\%, 62.0\%, 43.9\%, 52.4\%, {48.6\%}, 53.1\% and
{43.2\%} than {\tt DAX}, {\tt PMem}, {\tt NOVA}, {\tt BTT}, {\tt PMBD}, {{\tt PMBD-70}}, {\tt LRU} and {{\tt COA}} respectively, on Fileserver.
Fileserver is  write-intensive  and 
emulates I/O activities for a multi-threading 
file-server that
generates numerous write operations for multiple files.
\casaba is effectual because of concurrent eager evictions and flexible conditional bypasses.
As to Varmail
that simulates a 
mail-server and 
performs mandatory synchronization operations (i.e., {\tt fsync}) after writing files, 
\casaba takes 21.7\%, 22.4\%, 26.8\%, {31.8\%}, 8.2\% and {30.9\%} less time 
than {\tt NOVA}, {\tt BTT}, {\tt PMBD}, {{\tt PMBD-70}}, {\tt LRU} and {{\tt COA}}.
Though, it is slightly inferior to {\tt DAX} and {\tt PMem}
since the latter two have no cost of guaranteeing 
block-level atomic writes and flushing buffered data.

As for two read-intensive workloads, i.e., Webserver and Webproxy, 
\casaba still achieves 
comparable 
performance with other algorithms.
Webserver produces a sequence of read operations on files and simulates a  web-server's I/O activity,  
while Webproxy emulates  a   web-proxy server. Interestingly,
on Webproxy,
\casaba takes 12.7\%, 14.3\%, 11.9\%, {15.3\%}, 9.6\% and {13.4\%} less time 
than {\tt NOVA}, {\tt BTT}, {\tt PMBD}, {{\tt PMBD-70}}, {\tt LRU} and {{\tt COA}}. 
The reason is that Webproxy consists of a mix of write and read operations
while Webserver is dominated by read operations. 
{
	In particular,
	we collect the read latency with Filebench's workloads and present them 
	in\mbox{~\autoref{fig:qemu-read}}. It is evident
	that compared to other caching algorithms, \casaba's eager eviction hardly affects its read latency.
}

\section{{Related Work}}\label{sec:related}

{In this section, we discuss related works mainly on cache management  and storage
	systems built on PMem}.

\subsection{Cache Management }

{Caching has been proven to be useful for storage systems. 
	Researchers
	explored various novel cache management policies  tailored with regard to specific access patterns and the usage status of storage systems}~\cite{co-Active,PMem:PMBD:MSST-2014,FS:HiNFS:tos,cache:TreeFTL:TC-2016,SUN2023102896:DAC,TANG2023102930:SeqPack,SSD:DevFS:FAST-2018,ml:TCRM,ml:NCache}.

\textbf{SSD cache management.} 
{Leveraging the  device cache to buffer metadata for address translation or data
	has been thoroughly studied in the development of SSDs}~\cite{cache:DFTL:ASPLOS-2009,5496995,10.1145/2228360.2228518,10.1145/2947658}. 
For example, 
{For example, Wang and Wong}~\cite{cache:TreeFTL:TC-2016} {designed
	TreeFTL that organizes address translation and data pages in a tree-like structure in the device cache of SSD. 
	TreeFTL dynamically adjusts the partitions for address mapping and data buffering in order to adapt to a workload's access patterns.}
{Sun et al.}~\cite{SUN2023102896:DAC}  {proposed dynamic active and collaborative cache management, namely DAC,
	with a cache composed of cold/hot caches and ghost cold/hot caches. 
	DAC adjusts the real cache size based on observing cold/hot data in I/O requests with ghost caches.}
{The idea of \casaba is portable with SSDs,  especially NVMe ones 
	with fast access speed and multiple
	hardware queues}~\cite{10.5555/3358807.3358858,9373910,SSD:LODA:SOSP-2021,co-Active}.

\textbf{Cache management in file system.} {Researchers have considered effectual cache management
	in developing file systems for the evolving storage stack.
	HiNFS is a promising design that uses DRAM buffer to accelerate direct accesses with NVM in the
	file system}~\cite{FS:HiNFS:tos}. 
	In short, HiNFS employs a write buffer in DRAM to cache lazy-persistent writes while performing direct access to PMem for eager-persistent writes. 
{Kannan et al.}~\cite{SSD:DevFS:FAST-2018} {proposed 
	DevFS that is a device-level file system, providing performant direct-access capabilities within a storage device.
	In particular,
	DevFS employs reverse caching to move inactive  data structures of the file system
	off the device to host memory and coordinates with the OS to ensure secured file access.
	\casaba shares similarities with HiNFS and DevFS in caching data 
	by using a part of host OS's DRAM space,   instead of  a device's internal cache, for underlying storage. 
	Whereas, \casaba takes effects as I/O transit rather than I/O staging.}

\textbf{Machine learning-based cache management.}
{Recently, a few researchers have taken 
	machine learning (ML) based approaches for cache management}~\cite{9407137,ml:TCRM,ml:NCache}. 
For example, at the CPU cache level,
Sethumurugan et al.~\cite{9407137} {used machine learning as an {offline} tool to design a new replacement policy for CPU cache. 
	TCRM}~\cite{ml:TCRM} {addresses the trade-off between thermal and cache contention-induced slowdowns. It uses a neural network-based model to predict slowdown caused by cache contention. 
	At the storage cache level,
	NCache}~\cite{ml:NCache} {
	uses an ML model to predict data reaccess before eviction, preferentially evicting data unlikely to be accessed again and conserving cache space for frequently accessed data.
	The idea of \casaba and such ML-based techniques complement each other. 
	The way \casaba separates cold   from hot data through LRU evictions is simple but a bit coarse-grained.
	These ML-based techniques can help \casaba  to make a fine-grained separation.
	However, \casaba has to schedule an offline training and profiling, as it is non-trivial
	to build an ML framework in Linux kernel. Moreover, the workloads \casaba serves 
	vary significantly in different environments and over time. This is the second concrete challenge
	that \casaba shall  consider when employing ML-based techniques. We leave 
	this work for future exploration.}

{In all, these works aim to advance cache management policies, with a focus on improving  performance, reducing access latency, and enhancing load and store efficiency 
	in storage systems.
	Comparatively, \casaba boosts the performance of PMem-based block devices 
	by leveraging a DRAM buffer with I/O transit strategy
	to accelerate writes.}

\subsection{Storage Systems on PMem}

{The development of NVM technologies has motivated researchers to develop various
	approaches in effectively utilizing NVM as PMem}
~\cite{FS:NOVA:FAST-2016,CHEN201818:UMFS,YANG2022102629:VSM,PMSort,LI2022102547:MuHash,WANG2023102777:Reno}.

\textbf{File system on PMem.}
{Developing  new file systems for PMem has drawn wide attention, such as PMFS}~\cite{FS:PMFS:Eurosys-2014}, HiNFS, and NOVA.
	They mainly follow the DAX fashion that does not take PMem as  block device but memory.  
{Chen et al.}~\cite{CHEN201818:UMFS} {considered improving file access speed by minimizing kernel overhead. They expose files into user-space in constant time independent of file sizes. They also implement efficient user-space journaling for  consistency.
	Yang et al.}~\cite{YANG2022102629:VSM} {addressed problems caused by physical superpages in PMem file system. 
	Their design utilizes virtual superpages and includes Multi-grained Copy-on-Write (MCoW) and Zero-copy File Data Migration (ZFDM) mechanisms to reduce write amplification and improve space utilization efficiency.
	The concept of \casaba can be applied to DAX-based file systems developed for PMem. Although
	such file systems do not format PMem with BTT to be block devices,
	they mainly manage PMem space in the unit of pages and tend to perform data updates
	in the COW fashion for data consistency. These leave an opportunity for \casaba.}

\textbf{The use of DRAM-PMem system.}
{The slower access speed of PMem has motivated researchers to build a hybrid DRAM-PMem system}~\cite{PMSort,ZHAN201860,ZOU2022102462,WANG2023102777:Reno}.
{Sorting, indexing, and KV stores have been studied with DRAM-PMem.}
{Hua et al.}~\cite{PMSort} {conducted extensive experiments on various sorting methods adapted for PMem and further considered designing PMem-friendly sorting techniques on DRAM-PMem system. The 
	PMSort they proposed   adaptively  selects optimal algorithms and reduces failure recovery overhead.
	Li et al.}~\cite{LI2022102547:MuHash} {designed
	MuHash that is  a novel   persistent and concurrent hashing index for DRAM-PMem. 
	MuHash employs a multi-hash function scheme to solve the cascading write problem in PMem-based open-addressed hash-based indexes. 
	Wang et al}.~\cite{WANG2023102777:Reno} {presented a server-bypass architecture for KV stores on DRAM-PMem system. Their design 
	incorporates hopscotch hashing for latch-enabled append operations and a fully server-bypass read/write paradigm, so as  
	to eliminate network round trips and reduce hash conflicts for efficient client access to KV stores.}
{\casaba can also be viewed as a design built on DRAM-PMem system. However, \casaba works at a lower level as part of device driver. In addition,
	indexing, sorting, and KV stores need application-level caches}~\cite{LSM:AC-key-cache:FAST-2020,ZOU2022102462}, {which can be managed with \casaba's I/O transit
	strategy.}

\section{Conclusion}\label{sec:conclusion}

In this paper,
we revisit the use of PMem as block device and
consider adding a cache to BTT that achieves block-level write atomicity. 
We develop \casaba which, in contrast to 
I/O staging caches,
promptly transits buffered data into PMem in order to avoid
I/O stalls 
caused by full cache or {\tt fsync}s. To further alleviate I/O congestion,
it conditionally bypasses full cache without waiting for the drain of cache slots.
Leveraging multi-core CPU,
\casaba achieves high concurrency and scalability. It also retains block-level write atomicity.
We have thoroughly evaluated \casaba.
\casaba substantially boosts performance for BTT by as much as 3.6$\times$ in extensive experiments conducted
with both micro-benchmarks and real-world applications.

\section*{Acknowledgment}

This work was jointly supported by National Key R\&D Program of China No. 2022YFB4401700, Natural Science Foundation of Shanghai under Grants No. 22ZR1442000 and 23ZR1442300, and ShanghaiTech Startup Funding.
We are very grateful to Mr. Meng Chen for his valuable  help.

In addition, this paper was eventually accepted for production by the Journal of Systems Architecture: Embedded Software Design (JSA) 
\footnote{The website of JSA: \url{https://www.sciencedirect.com/journal/journal-of-systems-architecture}} after 10 months (12 May 2023 to 10 March 2024) of processing.
The Associate Editor of JSA who was in charge of this paper invited 
overall fourteen (14) reviewers and the paper underwent five (5) rounds of revisions.
To our best knowledge, such a number of reviewers should be very rare for a research paper  in the domain of electrical engineering and computer science.

\printcredits

\bibliographystyle{cas-model2-names}

\bibliography{io}

\subsection*{}
\setlength\intextsep{0pt}
\begin{wrapfigure}{l}{25mm}
	\centering
	\includegraphics[width=1in,height=1.25in,clip,keepaspectratio]{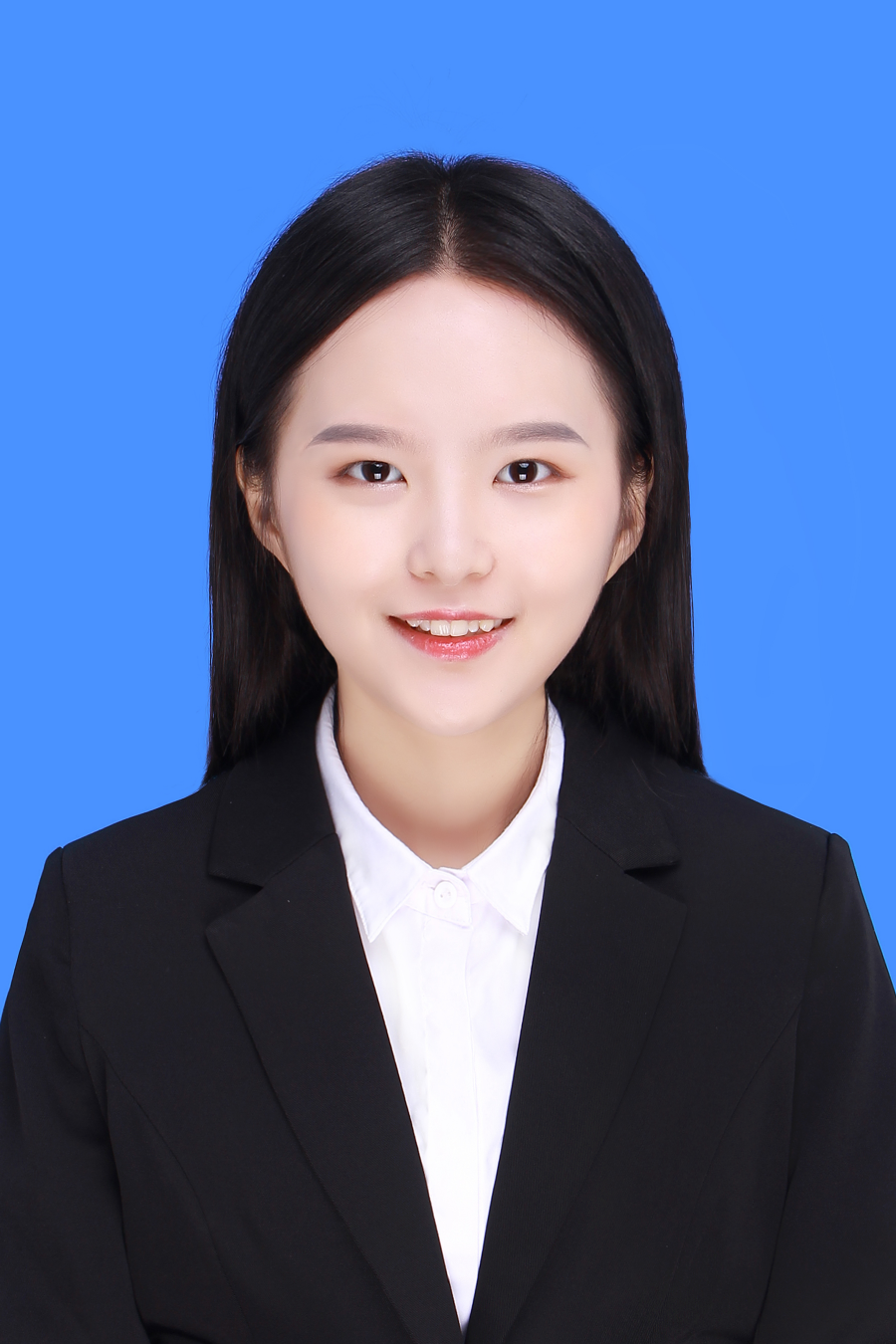}
\end{wrapfigure}
\noindent \textbf{Qing Xu} received her B.Eng. degree in software engineering from Hunan Normal University in 2021. She
is currently a Master's student majoring in computer science in ShanghaiTech University. Her research interests include file systems, data storage, and persistent memory.

\subsection*{}
\setlength\intextsep{0pt}
\begin{wrapfigure}{l}{25mm}
	\centering
	\includegraphics[width=1in,height=1.25in,clip,keepaspectratio]{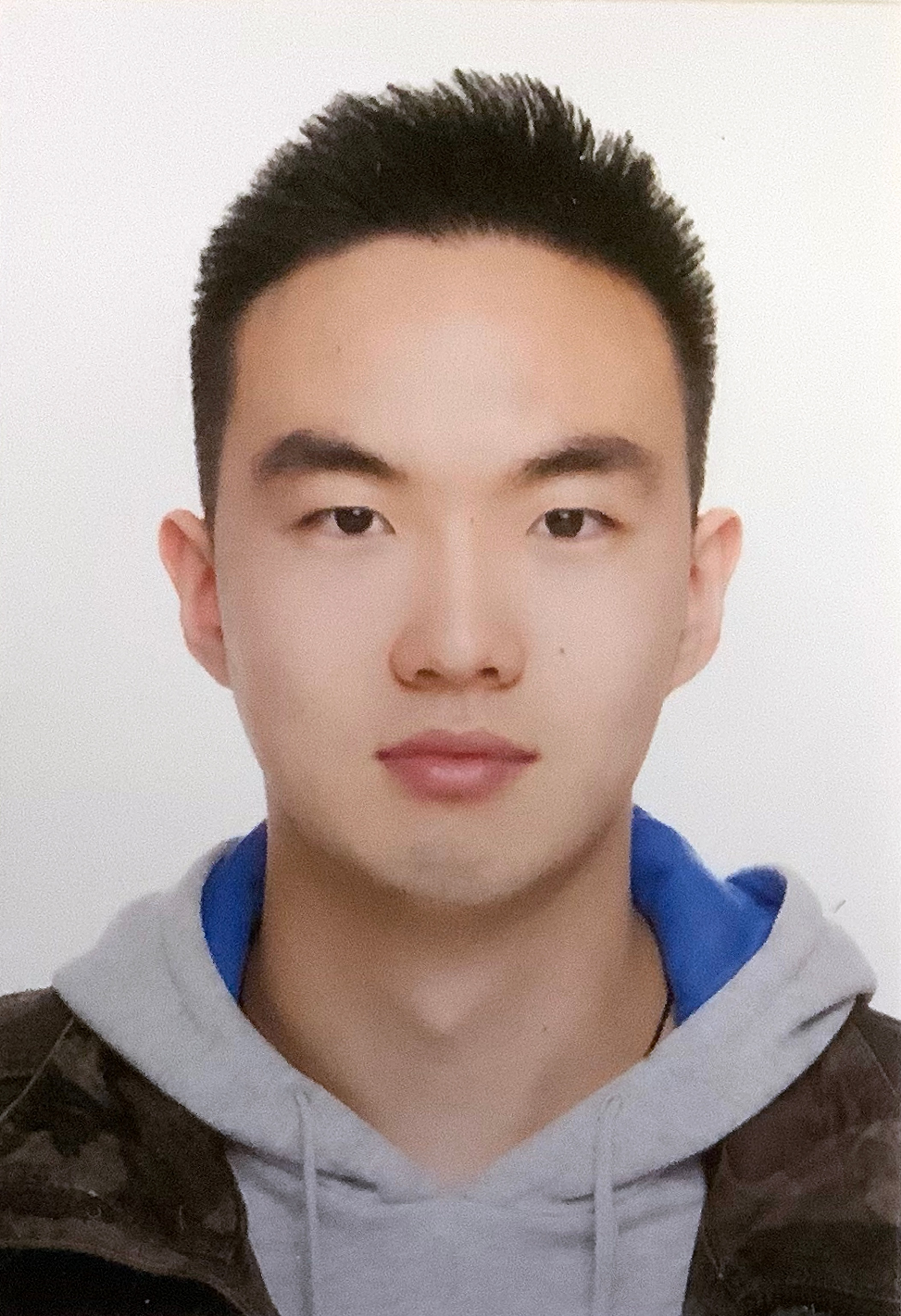}
\end{wrapfigure}
\noindent \textbf{Qisheng Jiang} obtained his B.Eng. degree in software engineering from Tongji University in 2021. He
is currently a Master's student majoring in computer science in ShanghaiTech University. Qisheng's research interests include systems for AI, persistent memory, and key-value store.

\subsection*{}
\setlength\intextsep{0pt}
\begin{wrapfigure}{l}{25mm}
	\centering
	\includegraphics[width=1in,height=1.25in,clip,keepaspectratio]{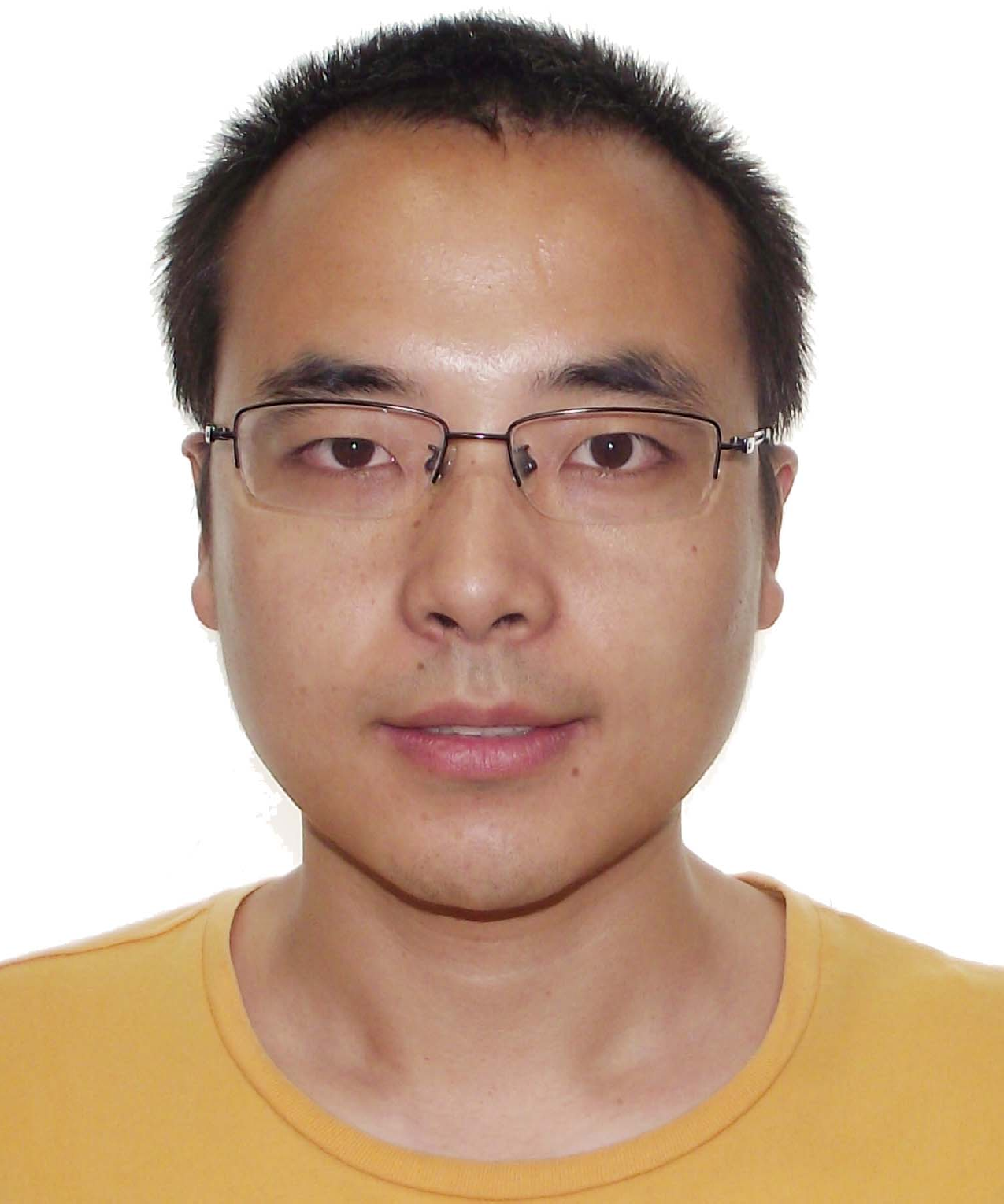}
\end{wrapfigure}
\noindent \textbf{Chundong Wang}
received the Bachelor's degree in computer science and technology from Xi'an Jiaotong University in 2008, and the Ph.D. degree in computer science
from National University of Singapore in 2013. Currently he works in ShanghaiTech University as a tenure-track assistant professor. Before joining
ShanghaiTech, he successively worked in Data Storage Institute, A$^\star$STAR, Singapore and Singapore University of Technology and Design (SUTD). 
He has published more than forty research papers in IEEE TC, IEEE TDSC, ACM TOS, ACM TECS, SC, DAC, USENIX Security, USENIX ATC, USENIX FAST, etc.
His research interests include data storage
and computer architecture.

\end{document}